\documentclass{aa}
\usepackage{graphicx}
\usepackage{xcolor}
\usepackage{siunitx}
\usepackage{txfonts}
\usepackage{natbib,twoopt}
\usepackage[breaklinks,unicode=true,colorlinks,linkcolor=red,citecolor=red,pdfdisplaydoctitle]{hyperref} 
\bibpunct{(}{)}{;}{a}{}{,}             
\makeatletter
  \newcommandtwoopt{\citeads}[3][][]{\href{http://adsabs.harvard.edu/abs/#3}%
    {\def\hyper@linkstart##1##2{}%
     \let\hyper@linkend\@empty\citealp[#1][#2]{#3}}}
  \newcommandtwoopt{\citepads}[3][][]{\href{http://adsabs.harvard.edu/abs/#3}%
    {\def\hyper@linkstart##1##2{}%
     \let\hyper@linkend\@empty\citep[#1][#2]{#3}}}
  \newcommandtwoopt{\citetads}[3][][]{\href{http://adsabs.harvard.edu/abs/#3}%
    {\def\hyper@linkstart##1##2{}%
     \let\hyper@linkend\@empty\citet[#1][#2]{#3}}}
  \newcommandtwoopt{\citeyearads}[3][][]%
    {\href{http://adsabs.harvard.edu/abs/#3}
    {\def\hyper@linkstart##1##2{}%
     \let\hyper@linkend\@empty\citeyear[#1][#2]{#3}}}
\makeatother
\begin{document} 

   \title{Mechanical strength distribution in Geminid meteoroids derived via fireball modeling}
   \titlerunning{Mechanical strength distribution of Geminids}
   \author{Tom\'a\v{s} Henych\inst{1}\fnmsep\thanks{Corresponding author, email: ftom@physics.muni.cz}
     \and
     Ji\v{r}\'i Borovi\v{c}ka\inst{1}
     \and
     Vlastimil Voj\'a\v{c}ek\inst{1}
     \and
     Pavel Spurn\'{y}\inst{1}
   }
   \institute{Astronomical Institute, Czech Academy of Sciences, Fri\v{c}ova 298, 251 65 Ond\v{r}ejov, Czech Republic}

   \date{Received 30 November 2023; accepted 25 January 2024}
 
   \abstract
   {Geminids are the most active annual meteor shower observed on Earth. Their parent is an active asteroid, (3200)~Phaethon, which is a target of the planned $\rm{DESTINY}^+$ mission of the Japan Aerospace Exploration Agency (JAXA). The exact physical nature of (3200)~Phaethon and Geminids is still debated.}
   {This paper is devoted to fragmentation modeling of bright Geminid fireballs, which should reveal information about the structure of centimeter-sized Geminid meteoroids. These fireballs were observed by the European Fireball Network (EN) over the past few years. We aim to describe their disintegration cascade in the atmosphere and their mechanical properties, and to derive their precise initial masses and velocities.}
   {We used a semi-empirical fragmentation model that employs an automatic procedure based on parallel genetic algorithms to determine the aerodynamic pressures at which a meteoroid and its parts fragment. This serves as a proxy for the mechanical strength of the body and its subsequent fragments. It enabled us to derive the minimum, median, and maximum mechanical strength and the strength distribution inside the meteoroid and reveal its internal structure.}
   {We find that the Geminids begin to crumble at pressures $1\rm{-}100\,\rm{kPa,}$ with the strongest parts reaching pressures of between $0.4$ and $1.55\,\rm{MPa}$ before fragmenting. Knowing the spectral type of (3200)~Phaethon (a B-type asteroid, part of the C complex), we conclude that the Geminids are made of compact and coherent carbonaceous material. We also find that the minimum aerodynamic pressure that causes the fragmentation of Geminids increases with increasing entry mass of Geminids. In contrast, the median aerodynamic pressure decreases as their entry mass increases. The spectra of all the observed Geminid fireballs show normal content and little variation in terms of sodium.}
   {}

   \maketitle

\section{Introduction}\label{intro} 
Geminids are the most active annual meteor shower; they are observed in the first half of December. Geminid meteoroids have atypical short-period orbits with high eccentricities and a very small perihelion distance of $\sim\!0.14\,\rm{AU}$ \citep{plavec1950}. While meteor showers are usually caused by comets, the Geminid orbits suggest an asteroidal origin. The material properties of Geminid meteoroids are therefore of high interest.

\citet{ceplecha1992} used their model of gross fragmentation for several best-observed Geminids, which allowed them to compare the photometric and dynamic mass of the meteoroids and thus derive their bulk densities. The calculation yielded values in the range $3000\rm{-}4000\,\rm{kg\,m^{-3}}$. High bulk densities indicated by a high ablation resistance were also suggested in \citet{spurny1993}. \citet{baba2002} obtained the value of $2900\pm600\,\rm{kg\,m^{-3}}$, assuming a process of quasi-continuous fragmentation. Using the equation of heat conductivity with assumed physical constants for meteoritic material and assuming quasi-continuous fragmentation, \citet{baba2009} derived both the grain and bulk densities of Geminids. These densities, $2900\pm200\,\rm{kg\,m^{-3}}$ and $2900\pm600\,\rm{kg\,m^{-3}}$, respectively, are significantly higher than for other meteor showers, implying negligible porosity. 

The mechanical strength of Geminids was previously derived by \citet{beech2002}, who assumed that the fast flickering observed in the light curves of some Geminids was caused by their fast spin. That led to rotational bursting rather than regular fragmentation caused by aerodynamic pressure. He derived the tensile strength of three Geminids to be roughly $3\cdot10^5\,\rm{Pa}$ for an assumed bulk density of $1000\,\rm{kg\,m^{-3}}$.

\citet{trigo2006} derived the mechanical strength proxy of small Geminids ($2.2\pm0.2\cdot10^4\,\rm{Pa}$) and meteoroids related to other meteor showers, assuming that the maximum brightness observed in these meteors was reached during the major breakup event and that this fragmentation was caused by mechanical forces. Several other values of the aerodynamic pressure that causes the major fragmentation of Geminids have been published, namely \citet{madiedo2013} derived a value of $3.8\pm0.4\,\rm{MPa}$ and \citet{prieto2013} derived $2.0\pm0.4\cdot10^4\,\rm{Pa}$ and $3.4\pm0.4\cdot10^4\,\rm{Pa}$ for two brightest flares observed in the light curve. \citet{neslusan2015} reviewed Geminid observations and their physical and chemical properties.

The parent body of Geminids is (3200)~Phaethon \citep[hereafter simply Phaethon;][]{whipple1983,hughes1983,davies1984,fox1984,williams1993}, a well-observed active Apollo asteroid \citep{jewitt2010,jewitt2013,li2013,hui2017}. It was recently found that Phaethon does not release dust but rather sodium during its perihelion passage \citep{hui2023,zhang2023}.

Later, other kilometer-sized Apollo asteroids were discovered, namely (155140)~2005~UD \citep{ohtsuka2006,jewitt2006} and (225416)~1999~YC \citep{ohtsuka2008,kasuga2008}, which form the Phaethon--Geminid Complex \citep{ohtsuka2006}. Their physical characteristics are consistent with those of Phaethon, but their mutual dynamical link is still debated \citep{ryabova2019,kasuga2022}. A recent review of the Phaethon--Geminid Complex is presented in \citet{kasuga2019}.

The origin of the Geminid stream and the physical nature of Phaethon are still debated. Since the discovery of Phaethon in 1983, many papers have speculated on the creation of the Geminids from Phaethon. \citet{hunt1986} assumed a collision between two rock-like bodies and were able to reproduce the correct cross-section of the stream but could not explain the distribution of aphelia of observed Geminids. \citet{jones1986} combined the cometary activity of Phaethon with gravitational perturbations and could roughly reconstruct the properties of the Geminid stream but did not consider their model to be a detailed description of the history of Geminids. \citet{belkovich1989} considered several models for the ejection of particles from the cometary nucleus. \citet{gustafson1989} integrated 20 orbits of Geminids backward in time as well as the orbit of Phaethon itself. He showed that Phaethon is indeed a parent of Geminids and that the cometary mechanism of their ejection was more plausible than other mechanisms. \citet{williams1993} confirmed the connection between Phaethon and Geminids but found that the actual mechanism of their creation was unclear. \citet{licandro2007} observed visible and near-infrared spectra of Phaethon and concluded that its spectral and dynamical properties suggest that Phaethon is probably an activated asteroid rather than an extinct comet.

\citet{ryabova2007,ryabova2016} developed a numerical model of the Geminid stream, assuming Phaethon experienced short-duration cometary activity  2000 years ago. Her models can reproduce the basic structure of the stream, but the model details are at odds with the observed activity profile. \citet{jones2016} assumed cometary activity according to Whipple's ejection model and compared their model results to the International Astronomical Union (IAU) photographic meteor database. They found that the ejection velocities were a factor of at least 3 too low and concluded that Whipple's model is not capable of explaining the origin of Geminids from Phaethon. \citet{hanus2018} derived the bulk density of Phaethon to be $1670\pm470\,\rm{kg\,m^{-3}}$ for $D=5.1\,\rm{km}$ (or $1480\pm420\,\rm{kg\,m^{-3}}$ assuming the $D=5.7\rm{-}5.8\,\rm{km}$ suggested by radar observations and the extrapolation of the polarimetric measurements). They concluded that the density was not consistent with a cometary origin for Phaethon. \citet{yu2019} proposed a new model based on the long-term sublimation of ice inside Phaethon and found that the production of meteoroids could have been sufficient in the past millennium to explain the Geminid stream. The idea of ice sublimation was already suggested in \citet{kasuga2009}, but it is not clear whether Phaethon contains any ice. 

\citet{nakano2020} suggested that Phaethon might shed mass due to deformation from its fast spin. For an assumed size of $6.25\,\rm{km}$ and a bulk density of $500\rm{-}1500\,\rm{kg\,m^{-3}}$, it might exceed the critical rotation period, and its structure might be sensitive to failure. This process could have led to the creation of the Geminid stream in the past and to the currently observed activity around perihelion. However, Phaethon's current activity is well described by the release of sodium \citep{zhang2023}.

Finally, \citet{cukier2023} created a dynamical model of the Geminid meteoroid stream calibrated with Earth-based measurements and compared it to the observation of a dust environment near the Sun by the Parker Solar Probe (PSP). They found that the PSP data were more consistent with their model of a catastrophic formation scenario and least consistent with a cometary formation mechanism.

The unclear formation scenario and the actual nature of Phaethon and the Geminids is the main motivation for the present study. Another motivation is the notion that small Geminids tend to be weaker than large ones \citep{borovicka2022b}. Therefore, we address the question of the physical and mechanical properties of the Geminid meteoroid material by modeling their fragmentation cascade in the atmosphere. From our detailed model, we calculated the aerodynamic pressure at fragmentation times, which serves as a proxy for the mechanical strength of the material. We compared these values to other observed fireballs that either produced meteorites or had otherwise firmly derived mechanical properties. We place constraints on the possible character and mechanical properties of the material that forms Geminids. Moreover, under reasonable assumptions, we derived the probable strength distribution inside Geminid meteoroids as well as the asteroidal meteoroids. That constrained the structure of these meteoroids.

The paper is organized as follows. In Sect.~\ref{data_mod} we describe how we obtained the data and how we reduced them. Section~\ref{met_prop} contains the procedure of deriving the physical properties of meteoroids, and in Sect.~\ref{results} we present our results, which we discuss in Sect.~\ref{discussion}. Conclusions of this study are given in Sect.~\ref{conclusions}.

\section{Data and their modeling}\label{data_mod} 

\subsection{Data}
We modeled nine Geminids and eight asteroidal fireballs observed by the European Fireball Network~\citep[EN;][]{spurny2007,spurny2017}. The instrumentation and data reduction are summarized in detail in \citet{borovicka2022a}. The most current state of the network was recently briefly described in \citet{henych2023}. For modeling, we used a high-time-resolution radiometric light curve, an all-sky image photometric light curve, and the dynamics of the foremost fragment. For some Geminids, quality spectra were taken by spectral all-sky cameras of the EN.

\subsection{Semiautomatic modeling procedure}\label{firmpik}
The physical model of meteoroid fragmentation used in this study was described in \citet{borovicka2020b}, and the semiautomatic method of modeling was described in \citet{henych2023}. Here we only briefly recap the fragmentation model characteristics and the optimization method. 

In the model, the meteoroid ablation and deceleration in the atmosphere were calculated. The main body fragmented at manually sought fragmentation times into several discrete fragments (gross fragmentation), into instantly released dust grains, causing a short and bright flare, or into an eroding fragment that released dust grains over a longer time, causing a gradual brightening. The individual fragments then ablated separately and could later fragment repeatedly (if they were not already eroding). The calculation stopped when the fragment was completely consumed by ablation or erosion, or if its velocity dropped below $2.5\,\rm{km\,s^{-1}}$. The total brightness of the meteor and the foremost fragment's length along the trajectory were calculated and compared to the data.

The optimization procedure called FirMpik is based on genetic algorithms and it starts with a population of random but physically plausible solutions and these are then evolved. The fitness function is constructed as the inverse value of the reduced $\chi^2$ sum of the model fit to the radiometric and photometric light curves and dynamical data of the foremost fragment.

Throughout the modeling we used slightly higher grain density of $\rho_{\rm grain}=3000\,\rm{kg\,m^{-3}}$ for Geminids \citep{ceplecha1992,baba2009} and $\rho_{\rm grain}=3500\,\rm{kg\,m^{-3}}$ for asteroidal fireballs, the product $\Gamma A=0.8$ and the ablation coefficient $\sigma=0.005\,\rm{kg\,MJ^{-1}}$. The luminous efficiency used in the model depends on the mass and the velocity of the meteoroid, we used the same function as \citet{borovicka2020b}. For our calculations, we used densities from the NRLMSISE-00 atmosphere model \citep{picone2002}. In modeling of Geminids, we had to also use different dataset weights than we used in asteroidal fireballs modeling described in \citet{henych2023}. The basic optimization algorithm was the same as in the previous paper \citep{henych2023}, but we added a few new features that could lead to better results.

\subsubsection{Fragmentation time optimization}
The new feature of the FirMpik program is the automatic optimization of the fragmentation time. So far, we have searched for the fragmentation times in the radiometric curve manually. Two types of fragmentation times manifest themselves in the radiometric curve in different ways. The first appears as a bright and usually brief flare, and is both symmetric and asymmetric. This is likely a manifestation of a sudden dust release from the meteoroid and this type can be found in the data very precisely and therefore does not need any optimization whatsoever. The shape of this flare is dictated by the mass distribution of released dust particles.

The other type of fragmentation is related to a gradual erosion of dust particles from the meteoroid and it usually appears as a hump or a change of the slope of the radiometric curve. It is more difficult and less certain to find these fragmentation times and therefore we decided to optimize them in the process of finding a solution. The procedure is as follows: first, we find the most probable fragmentation times manually led by our previous experience from the modeling. Then, we decided which fragmentation times we fixed and which should be optimized (floating). We can set a maximum time shift for specific fragmentation times, but there is a limit to how much we can shift them. It prevents the floating fragmentation times from interfering with the previous or the next fragmentation time (whether fixed or also floating). This procedure increases the number of free parameters depending on the number of floating fragmentation times, but it may enable fitting the measured data with a higher accuracy and also to find the fragmentation times, heights, and in turn the aerodynamic pressure exerted on the meteoroid front surface more confidently.

\subsubsection{Forced erosion}
We forced erosion for some fragmentation times in scenarios where we expected the creation of some eroding fragment. We could also fix the erosion coefficient, $\eta$, and the mass distribution indexes that were found in previous runs. In these runs, we were able to fit certain features of the radiometric curve well, and we think the best solution will contain the eroding fragment with these properties as well. This way, we could make use of information gained in otherwise unsuccessful runs.

\subsubsection{Fragment multiplicity}
Multiple fragments are identical fragments created at the same fragmentation time \citep{borovicka2020b}. It helps to fit the very end of the radiometric light curve when the overall brightness drops fast after a bright flare. If a specific fragment is broken apart into several smaller fragments, their larger total cross-section causes a brightening. The masses of individual fragments are small and they all are consumed by ablation faster than a single fragment whose mass is the sum of the masses of these smaller fragments. This causes a faster drop in luminosity. It also saves some computation time because the calculation is done only for one fragment and then the luminosity is multiplied by the number of fragments.

\section{Meteoroid properties}\label{met_prop} 

\subsection{Mechanical strength}\label{mech_strength}
The aerodynamic pressure exerted on a meteoroid and its fragments just before they fragment is a proxy for their mechanical strength. The aerodynamic pressure is calculated as

\begin{equation}
  p = \Gamma\rho_{\rm a} v^2=\frac{C_{\rm D}}{2}\rho_{\rm a} v^2,
\end{equation}

\noindent where $\rho_{\rm a}$ is the density of the atmosphere at the fragmentation height, $v$ is the meteoroid or fragment velocity at that height and $C_{\rm D}$ is the drag coefficient. \footnote{The referee of the paper recommended against calling $\Gamma$ the "drag coefficient" to avoid confusing any aerodynamicist who might read this and other papers on meteors. We followed this advice in contrast to traditional meteoric literature, such as \cite{bronshten1983}, who was aware of the classical definition of the drag coefficient ($C_{\rm D}= 2\Gamma$).} We used $\Gamma=1$ ($C_{\rm D}=2$) in all our plots to be able to compare various meteoroids, but the actual value of $\Gamma$ used in our modeling is different. The model contains only the product of $\Gamma A$, where $A$ is the shape coefficient. In our modeling, $\Gamma A=0.8$ \citep{borovicka2020a}, so if we assume spherical fragments (a very common but obviously simplified assumption), $A\doteq1.21$ and $\Gamma\doteq0.66$. This value is consistent with a value used in \citet{ceplecha1998}. Should the meteoroids have a more aerodynamic shape and orientation, the product $\Gamma A$ would be as low as $0.2$. Conversely, a flat tile with diameter to height ratio of 5:1 moving with the flat side down would have $A=2.9$ and $\Gamma=0.8$ \citep{zhdan2007}.

The unknown shape of the fragmenting meteoroid contributes the most to the aerodynamic pressure uncertainty. The random errors caused by the precision of velocity and density of the atmosphere are on the order of a few per cent for well-observed fireballs.

It should be noted that the aerodynamic pressure acting on the front part of the meteoroid should not be confused with its internal mechanical strength. A stress analysis of the loading process is required to derive the actual tensile strength (or any other appropriate strength measure) of the meteoroid. A similar calculation is done in structural mechanics for the design of buildings, bridges, and airplanes so that they can withstand the respective loading. This was done before for large meteoroids, asteroids, or comets \citep[e.g.,][]{shuvalov2002,boslough2008,robertson2017} by means of hydrocode simulations. Nevertheless, simulations for smaller meteoroids of various shapes, sizes, velocities, impact parameters, internal structure, and material properties are still lacking.

\citet{fadeenko1967} calculated the elastic stress field inside a (homogeneous) elastic spherical meteoroid and derived an analytical condition for its disintegration. He stated that for most materials the ultimate shear strength is much lower than the ultimate tensile strength and is only a weak function of the meteoroid mass. Therefore, we should instead observe the maximum shear stress condition, which is approximately

\begin{equation}
  \sigma_{\rm shear}\simeq0.265\rho_{\rm a} v^2,
\end{equation}rather than the maximum tensile stress condition,

\begin{equation}
  \sigma_{\rm tensile}\simeq0.365\rho_{\rm a} v^2.
\end{equation}

To the contrary, \citet{slyuta2017} claims that the shear stress distribution calculated by \citet{fadeenko1967} implies the ruling material strength is the tensile strength. This is also a prevailing assertion in several papers that dealt with deriving the overall mechanical strength of meteoroids. We conservatively endorse this statement, but we also encourage further research of the detailed failure mechanism of fragmenting meteoroids by means of proper stress analysis.

Fragment masses and the aerodynamic pressure at which they split are used to construct a plot in which we compare Geminids, asteroidal fireballs, the Winchcombe fireball (which dropped CM2 meteorites), and Taurids. That plot enables us to place constraints on the mechanical strength proxy and the corresponding composition of Geminid meteoroids.

The value of aerodynamic pressure can also help us inspect the internal structure of a meteoroid. We used the following processes to derive a mass fraction of the entry mass of the meteoroid (the initial meteoroid body is called the Main in our terminology) that is destroyed at a specific aerodynamic pressure value.

Dust grains are released immediately at a specific aerodynamic pressure when a gross fragmentation occurs. The meteoroid is usually split into several macroscopic fragments, and some amount of dust is released.

A gradual erosion of dust grains also begins at a distinct aerodynamic pressure value. There are two possibilities. We can assign the eroded grains to a fragmentation pressure if the eroding fragment is broken apart at an instant of fragmentation to discrete dust grains that are later released and ablated. Or we can assign parts of the eroding fragment (in discrete timesteps of our model) to different pressure values if the bits and grains are gradually released from an eroding fragment as the aerodynamic pressure grows when the eroding fragment travels deeper into the atmosphere. We are more inclined to the former possibility and the plots in Sect.~\ref{results} are constructed accordingly. Nevertheless, we cannot exclude the latter possibility. It is conceivable that for faster erosion, the aerodynamic pressure that breaks apart the parent fragment is a more appropriate value. For slower, more gradual erosion lasting several seconds (comparable in duration to the whole fireball event), it would be more appropriate to choose the latter calculation method.

For non-eroding and non-fragmenting macroscopic fragments we can at least set a lower limit of an aerodynamic pressure that did not break the fragment. This is the maximum pressure reached by the specific fragment on its travel through the atmosphere and the actual mechanical strength of the fragment is higher than this value. 

The mass fraction of the entry mass of the meteoroid destroyed at a specific aerodynamic pressure value is calculated as the ratio of the dust mass (or an eroding or non-fragmenting fragment mass) to the mass of its parent fragment. This ratio is then multiplied by the fraction of the mass of the parent fragment to the mass of its parent fragment and so on until we find the fraction of the mass in question to the Main mass. All this is expressed in Eq.~\ref{dust_rat} for dust,

\begin{equation}
m\!f\!d_{\rm x}=\frac{m_{\rm dust_x}}{m^{\rm f}_{\rm x}}\prod^{\rm Main}_{\rm x=fragment}\frac{m^{\rm i}_x}{m^{\rm f}_{x-1}},
\label{dust_rat}
\end{equation}and~Eq.~\ref{ero_rat} for an eroding or non-fragmenting fragment,
\begin{equation}
m\!f\!\!f_{\rm x}=\prod^{\rm Main}_{\rm x=fragment}\frac{m^{\rm i}_x}{m^{\rm f}_{x-1}},
\label{ero_rat}
\end{equation}where $m\!f\!d$ and $m\!f\!\!f$ are the mass fractions, $m$ are masses, the superscript~i indicates the initial mass (after fragment creation), and the superscript~f indicates the final mass (before the fragmentation). We note that this calculation is an approximation that includes also the meteoroid part that ablates and whose mechanical properties cannot be directly derived.

\begin{figure}
   \centering
     \includegraphics[width=\hsize]{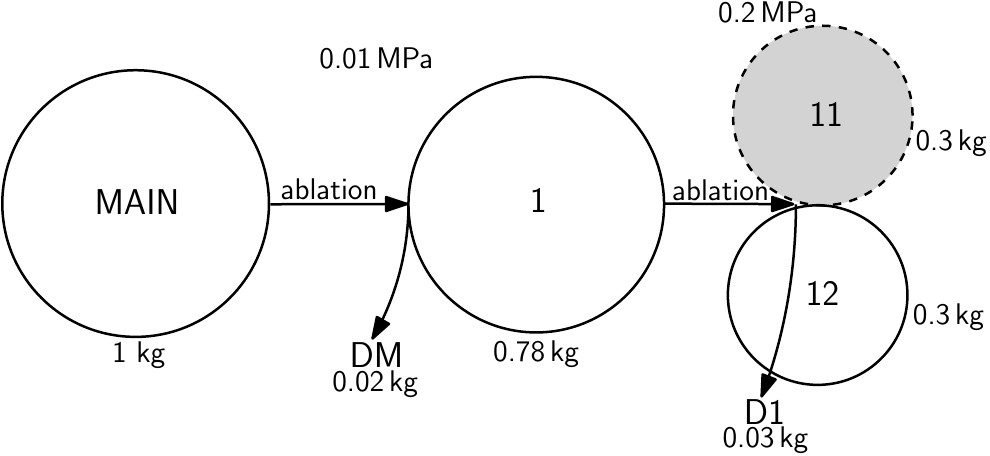}
      \caption{Cartoon describing the calculation of the meteoroid portion destroyed at a specific aerodynamic pressure. This pressure is the proxy for its mechanical strength. Fragment names are inside the circles, and shaded fragment 11 (with a dashed border) is eroding. The dust name starts with a D.}
      \label{fraction_cartoon}
\end{figure}

Figure~\ref{fraction_cartoon} shows an example of such a calculation. The fraction of dust (DM) is simply

\begin{equation}
m\!f\!d_{\rm DM}=\frac{m_{\rm dust_{DM}}}{m^{\rm f}_{\rm Main}}=\frac{0.02\,\rm{kg}}{(0.78+0.02)\,\rm{kg}}=0.025,
\label{dm_expl}
\end{equation}and this fraction of the entry mass was destroyed by the aerodynamic pressure of $0.01\,\rm{MPa}$. The fraction of the eroding fragment called 11 is

\begin{equation}
m\!f\!\!f_{11}=\frac{m^{\rm i}_{11}}{m^{\rm f}_1}\cdot\frac{m^{\rm i}_1}{m^{\rm f}_{\rm Main}}=\frac{0.3\,\rm{kg}}{0.63\,\rm{kg}}\cdot\frac{0.78\,\rm{kg}}{0.8\,\rm{kg}}=0.464.
\label{ero_expl}
\end{equation}Together with the dust D1 mass fraction (0.046), the total of 0.511 of the entry mass has the strength proxy of $0.2\,\rm{MPa}$.

To recap, the dust released in gross fragmentation is thought to be crushed by the aerodynamic pressure that caused the gross fragmentation. Eroding fragments have also the same strength proxy as the aerodynamic pressure fragmenting their parent fragment, the dust grains are thought to be created at the fragmentation time (and pressure) and then gradually released and ablated.

We can set a lower limit of the strength for non-eroding fragments that do not further fragment. It is the maximum aerodynamic pressure reached in their flight through the atmosphere before they decelerate and the pressure starts to decrease. Their mechanical strength must be higher than that pressure. In the above example, if fragment 12 did not further fragment and reached a maximum aerodynamic pressure of, say, $0.6\,\rm{MPa}$, the aerodynamic pressures that destroyed the original meteoroid were distributed as follows: $2.5\%$ of $0.01\,\rm{MPa}$, $51.1\%$ of $0.2\,\rm{MPa}$, and $46.4\%$ of $0.6\,\rm{MPa}$.

Some parts of asteroidal fireballs fall to the ground as meteorites. In principle, their strength could be measured should they be found. In Geminids, the production of meteorites is not very likely or extremely rare \citep[][Spurn\'y \& Borovi\v{c}ka, in~prep.]{beech2003,madiedo2013}.

This procedure is used to construct mechanical strength proxy distribution within the meteoroid. It is presented in Sect.~\ref{results} as a histogram for each modeled Geminid and asteroidal fireball, where the fragmentation aerodynamic pressure is binned in predefined bins. It is also presented as a pie chart for which we used the actual aerodynamic pressure values, not the binned values.

\subsection{Mass loss modes}\label{massloss}
The model enables us to discern various processes in which the mass of the meteoroid is lost. There are three such processes: an immediate dust release during a gross fragmentation, a regular ablation of fragments, and an erosion of dust grains. When plotted as a function of the entry meteoroid mass, we expect to see some trends as was the case in Taurids~\citep[see Fig.~8 in][]{borovicka2020a}.

\subsection{Mass evolution}\label{massevo}
It is interesting to plot the evolution of the total meteoroid mass as a function of aerodynamic pressure that is exerted on its front part. It includes the whole mass of the meteoroid in a specific timestep of the model, all the fragments, and dust grains. Then we can choose some threshold value of the total mass and compare the pressures at which all modeled Geminids reached the threshold value. In asteroid science, the threshold when the asteroid loses half of its original mass in a collision~\citep{fujiwara1989} or by other means~\citep{richardson1998} is called a catastrophic disruption and we think it is a useful criterion in meteoroid science, too. Therefore, we define a catastrophic disruption of a meteoroid as the moment when it loses half of its entry mass.

After fragmentation of the meteoroid, the fragments have various sizes and they spread along the flight trajectory. They are at different heights above the ground where the atmosphere has a different density and they also have different velocities because their decelerations differ. Therefore, the aerodynamic pressure exerted on these fragments is different, too. It is then necessary to plot some specific value of the aerodynamic pressure or the whole span of pressures.

\section{Results}\label{results} 
\begin{figure}
   \centering
     \includegraphics[width=\hsize]{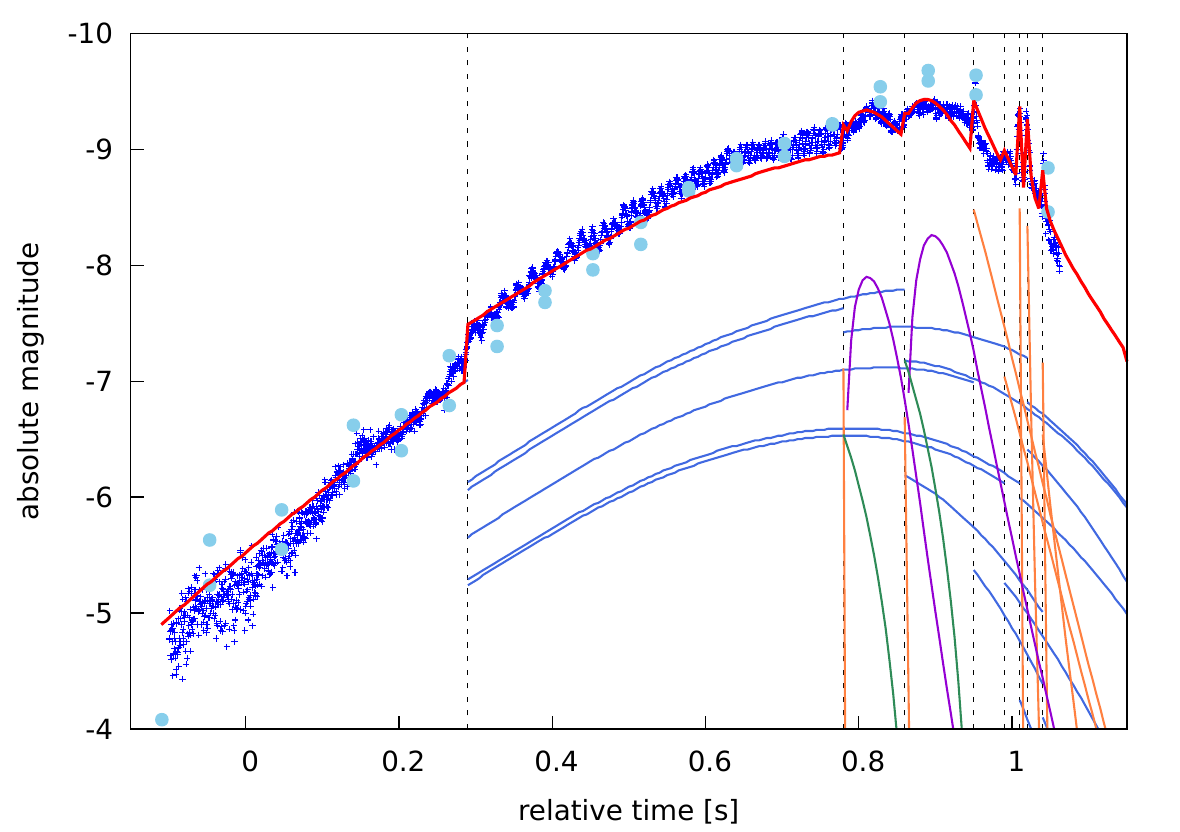}
      \caption{Automatic solution of the EN131218 fragmentation compared to the observed radiometric curve (dark blue pluses) and a photometric light curve (sky blue disks). The total model brightness is shown as a solid red line, the brightness of regular fragments is shown as blue curves, green curves signify eroding fragments, violet lines indicate dust particles released from these fragments, and orange curves denote regular dust released in gross fragmentations. Fragmentation times are shown with vertical dashed lines.}
      \label{gem6_rlc}
\end{figure}

\begin{figure}
   \centering
     \includegraphics[width=\hsize]{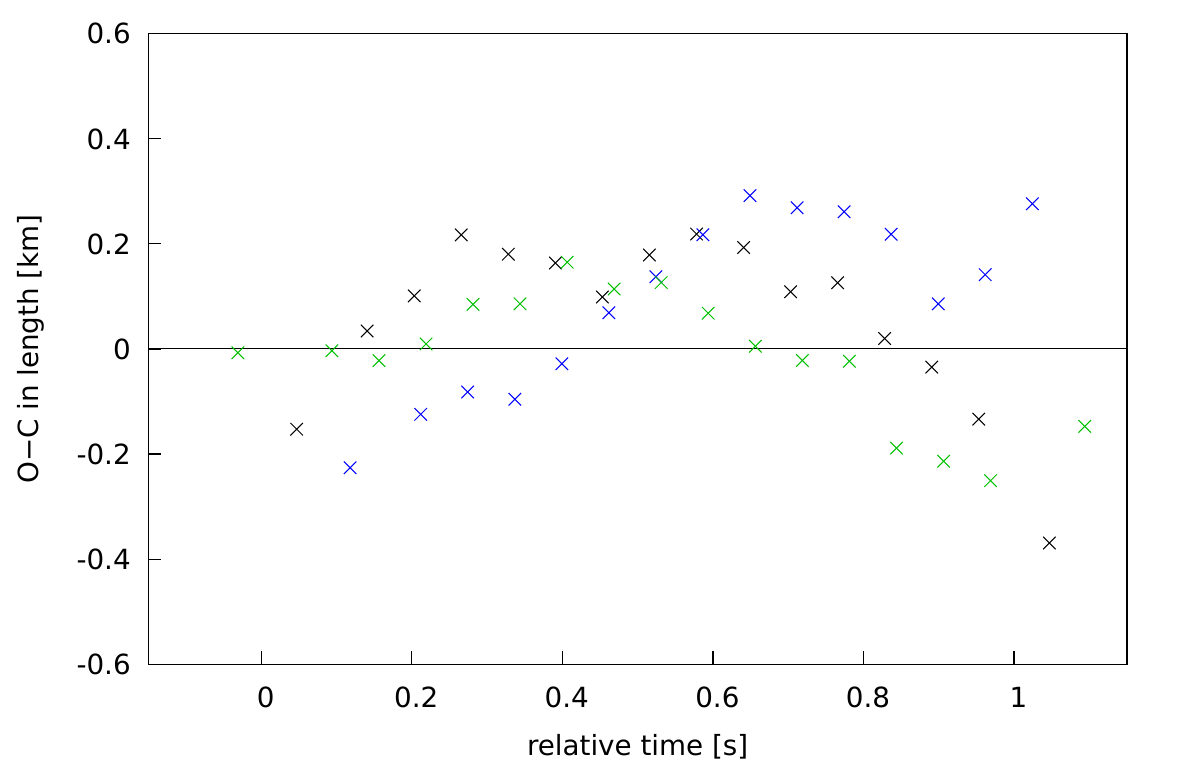}
      \caption{Residuals of length for an automatic solution of the EN131218. Different colors indicate the stations of the EN that were used to calculate the model.}
      \label{gem6_lenoc}
\end{figure}

The modeled Geminids were observed between 2018 and 2022. Their entry masses span almost two orders of magnitude, from $0.027\,\rm{kg}$ to $1.7\,\rm{kg}$. Their identification and physical properties are given in Table~\ref{gem_tab}. Figures~\ref{gem6_rlc} and~\ref{gem6_lenoc} show one example of a model found by the semiautomatic procedure described in Sect.~\ref{firmpik}. Figure~\ref{gem6_rlc} shows a radiometric light curve of the EN131218$\_$012640 (hereafter EN131218) fitted by the physical model and Fig.~\ref{gem6_lenoc} shows residuals of its length along the trajectory fit. The model fits the data very well.

We also modeled asteroidal fireballs to compare their properties with Geminids. There were three 
fireballs with masses between $0.066\,\rm{kg}$ and $0.19\,\rm{kg}$ and they were supplemented by five fireballs modeled in \citet{borovicka2020b} and \citet{henych2023} with entry masses between $2.4\,\rm{kg}$ and $6.5\,\rm{kg}$. We used the semiautomatic solutions from \citet{henych2023}  in this study. The physical properties of these fireballs are described in Table~\ref{oc_tab}.

\begin{table*}
\caption{Geminid fireballs that were modeled in the study, their entry mass, initial velocity, estimated diameter ($\rho_{\rm bulk}=2500\,\rm{kg\,m^{-3}}$), trajectory slope to the ground, and maximum magnitude.}
\label{gem_tab}
\centering
\begin{tabular}{c c S[table-format=2.1] c S[table-format=2.1] c S[table-format=2.1]}
\hline\hline
no. & fireball code & {mass [kg]} & velocity [km/s] & {diameter [cm]} & trajectory slope [$^\circ$] & {maximum magnitude} \\
\hline
g1 & EN151220$\_$024219 & 0.83 & 35.23 & 8.6 & 62.9 & -11.9 \\
g2 & EN141221$\_$194403 & 0.13 & 36.15 & 4.7 & 35.3 & -10.1 \\
g3 & EN131221$\_$024216 & 0.029 & 35.68 & 2.8 & 63.2 & -8.0 \\
g4 & EN141220$\_$185349 & 0.027 & 36.11 & 2.7 & 25.0 & -7.9 \\
g5 & EN121218$\_$193710 & 1.7 & 35.60 & 10.8 & 28.8 & -11.5 \\
g6 & EN131218$\_$012640 & 0.13 & 35.03 & 4.6 & 71.8 & -9.6 \\
g7 & EN091220$\_$210049 & 0.033 & 35.53 & 2.9 & 44.4 & -8.0 \\
g8 & EN141219$\_$200732 & 0.17 & 35.56 & 5.0 & 37.0 & -9.3 \\
g9 & EN141222$\_$220653 & 0.29 & 36.14 & 6.0 & 52.8 & -10.7 \\
\hline
\end{tabular}
\end{table*}

\begin{table*}
\caption{Asteroidal fireballs that were partly modeled in the study and partly taken from \citet{henych2023}, along with their entry mass, initial velocity, estimated diameter ($\rho_{\rm bulk}=3500\,\rm{kg\,m^{-3}}$), trajectory slope to the ground, and maximum magnitude.}
\label{oc_tab}
\centering
\begin{tabular}{c c S[table-format=2.1] c S[table-format=2.1] c S[table-format=2.1]}
\hline\hline
no. & fireball code & {mass [kg]} & velocity [km/s] & {diameter [cm]} & trajectory slope [$^\circ$] & {maximum magnitude} \\
\hline
a1 & EN011021$\_$234158 & 0.066 & 33.17 & 3.3 & 52.9 & -9.8 \\
a2 & EN101119$\_$190355 & 0.096 & 24.91 & 3.7 & 35.3 & -7.6 \\
a3 & EN220420$\_$232942 & 0.19 & 32.61 & 4.7 & 33.6 & -11.0 \\
a4 & EN240217$\_$190640 & 2.4 & 17.80 & 11.0 & 44.5 & -9.6 \\
a5 & EN180118$\_$182623 & 2.9 & 20.07 & 11.6 & 46.8 & -10.2 \\
a6 & EN020615$\_$215119 & 5.1 & 16.23 & 14.0 & 69.6 & -9.6 \\
a7 & EN110918$\_$214648 & 6.4 & 23.66 & 15.2 & 62.8 & -14.0 \\
a8 & EN260815$\_$233145 & 6.5 & 19.35 & 15.3 & 36.5 & -11.7 \\
\hline
\end{tabular}
\end{table*}

\subsection{Pressure--mass plot}\label{presmas}
The aerodynamic pressure exerted on a meteoroid and its fragments just before they fragment is a proxy for their mechanical strength. All these fragments are plotted in mass--pressure plot and compared to the asteroidal meteoroids, to the Winchcombe carbonaceous CM2 chondrite \citep{mcmullan2023}, and also to Taurids \citep{borovicka2020a}. All these fireballs were modeled with the same fragmentation model.

\begin{figure}
   \centering
   \includegraphics[width=\hsize]{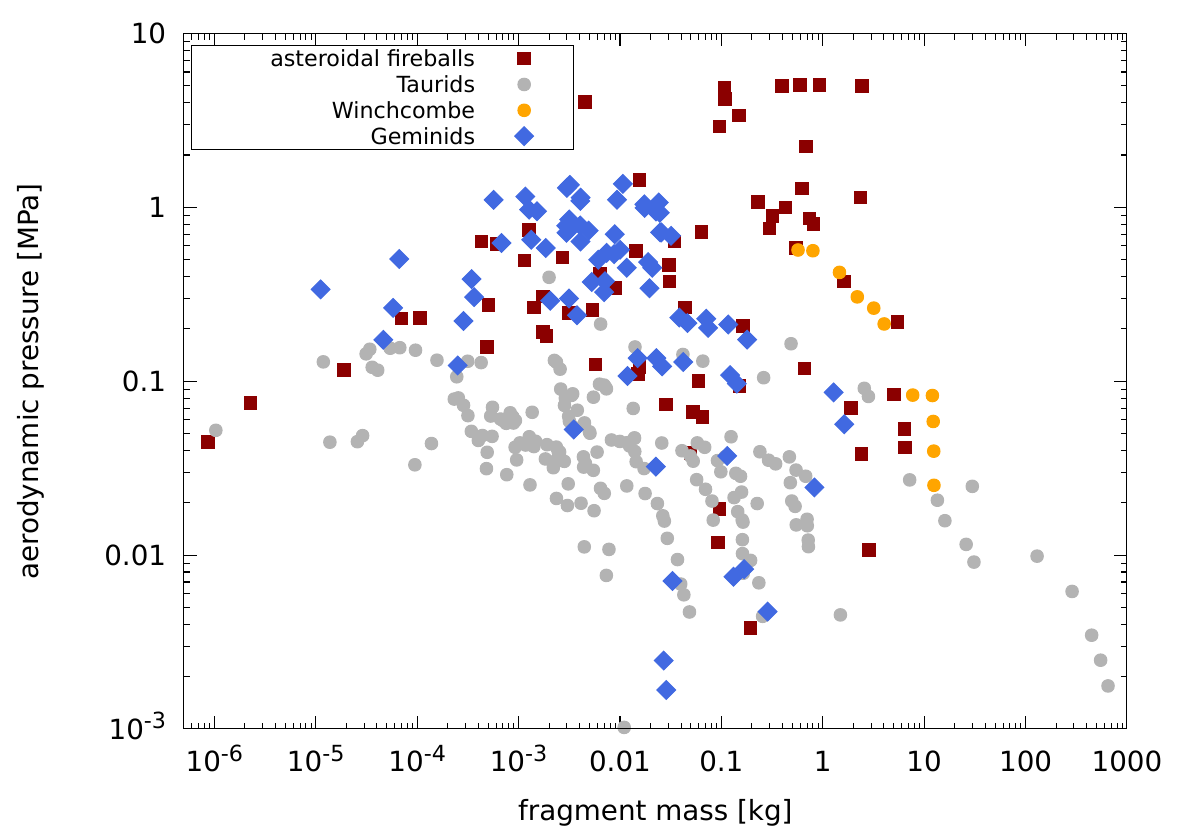}
      \caption{Geminid fragment mass distribution vs. aerodynamic pressure compared to asteroidal fireballs, Taurids, and the Winchcombe fireball that produced CM2 meteorites.}
      \label{pres_mass_gem}
\end{figure}

\begin{figure}
   \centering
   \includegraphics[width=\hsize]{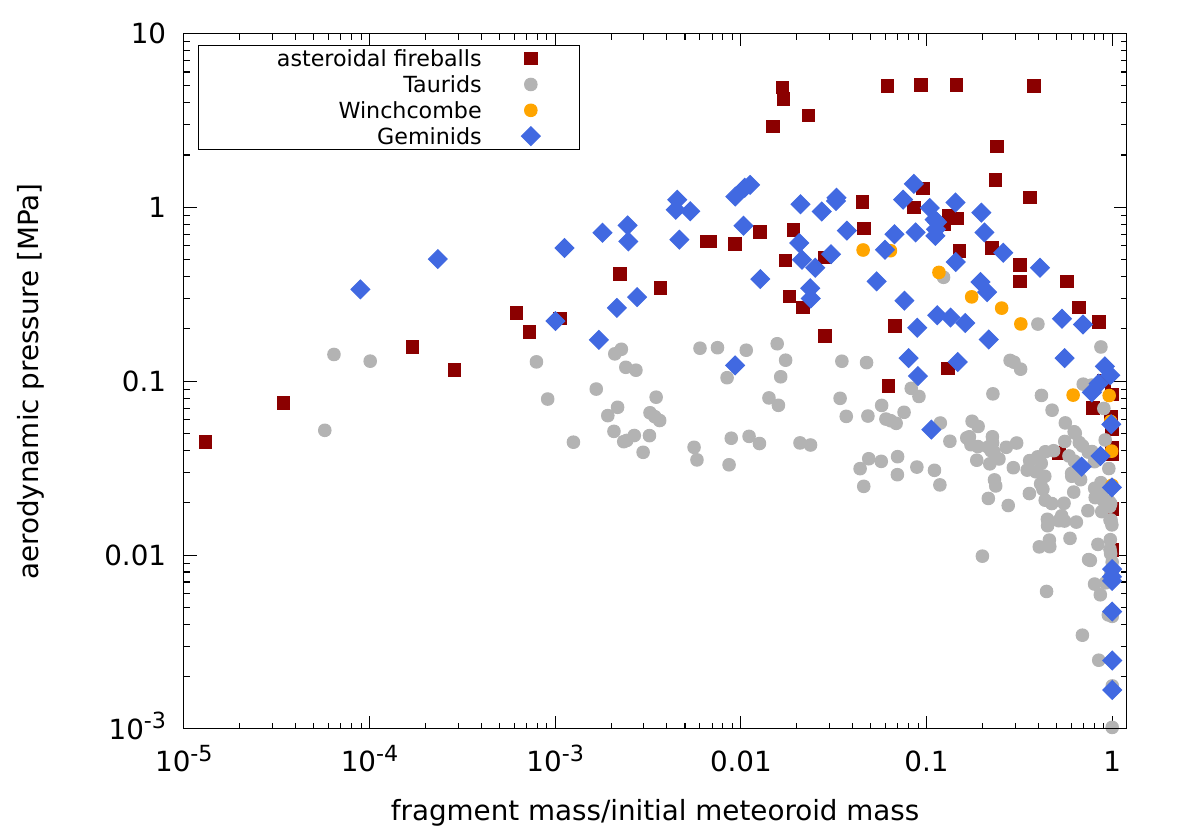}
      \caption{Geminid fragment mass distribution normalized by the initial mass of the meteoroid vs. aerodynamic pressure compared to asteroidal fireballs, Taurids, and the Winchcombe fireball that produced CM2 meteorites.}
      \label{pres_rel_mass_gem}
\end{figure}

Figures~\ref{pres_mass_gem} and~\ref{pres_rel_mass_gem} show that the first fragmentation occurs at a range of pressures from as low as $1\,\rm{kPa}$ to $0.1\,\rm{MPa}$. The maximum aerodynamic pressures at which the Geminids fragment are between $0.4\,\rm{MPa}$ and $1.55\,\rm{MPa}$ (Geminid~6), and the absolute maximum is about $3.4$ times higher than in Taurids. It is a well-known fact that Geminids are made of much stronger material than other shower meteoroids \citep{spurny1993}. It is also $2.4$ times higher than the maximum pressure reached in the Winchcombe meteoroid fragmentation, but $3.7$ lower than the maximum aerodynamic pressure reached by asteroidal fireballs of similar entry mass that we modeled for comparison. Further analysis of the mechanical strength distribution of Geminid meteoroids is presented in the next section.

\subsection{Strength distribution}
\begin{figure}
   \centering
   \includegraphics[width=\hsize]{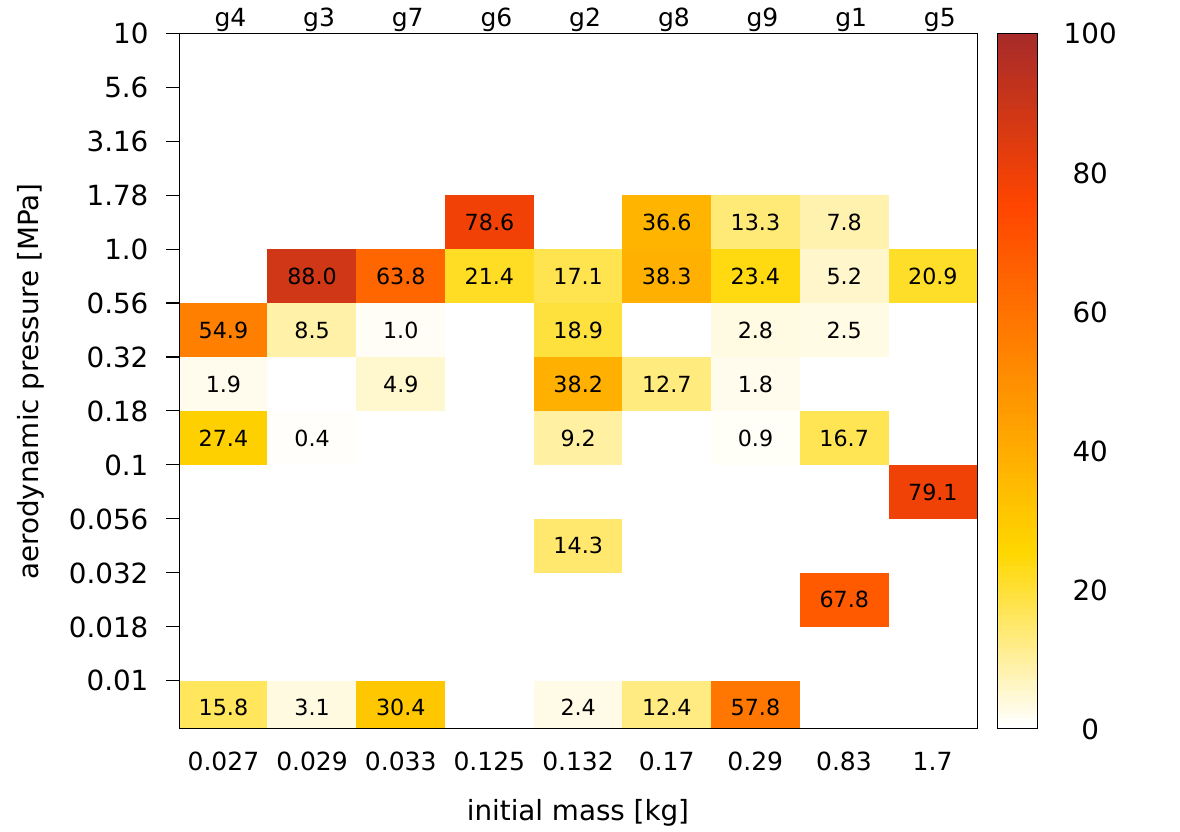}
      \caption{Aerodynamic pressure distribution vs. the initial mass of the meteoroid for Geminid fireballs in predefined bins. Numbers and colors designate the percentage of the entry mass destroyed at a certain aerodynamic pressure. See Table~\ref{gem_tab} for Geminid identifications.}
      \label{stren_dist_gem}
\end{figure}

\begin{figure}
   \centering
   \includegraphics[width=\hsize]{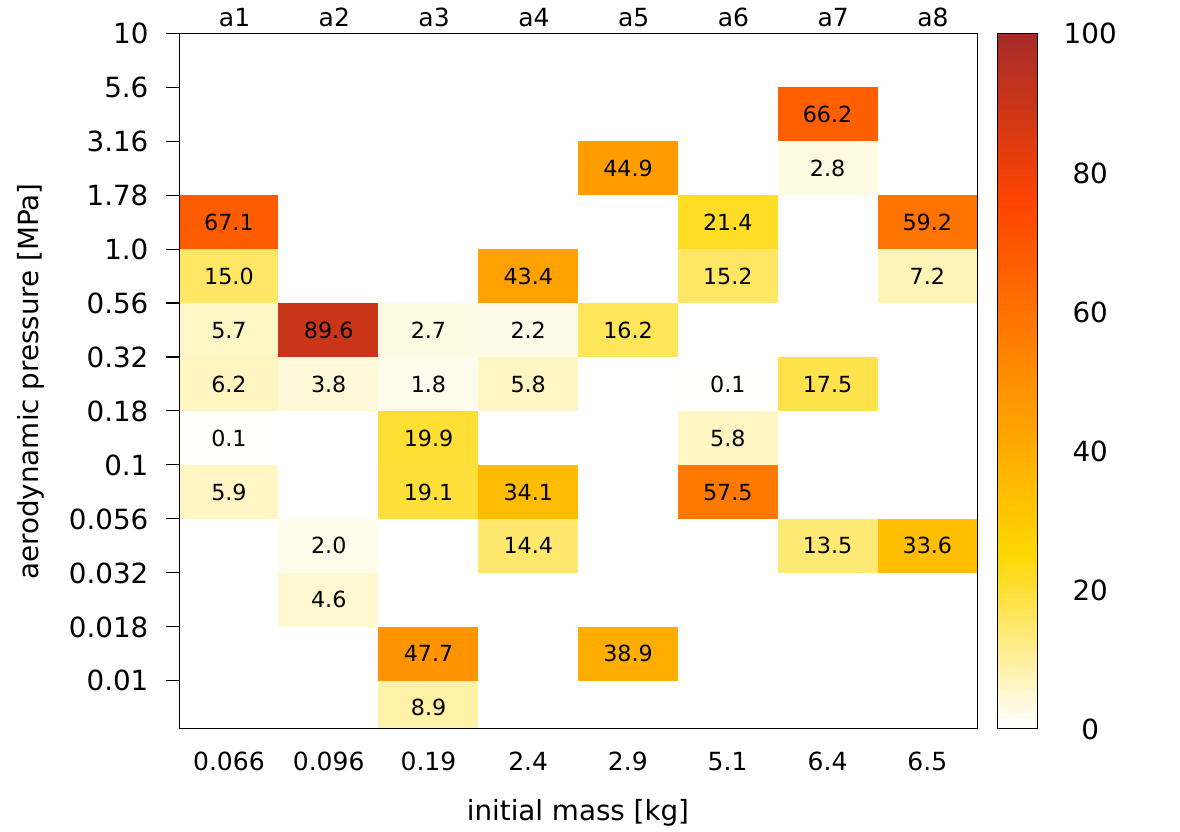}
      \caption{Aerodynamic pressure distribution vs. the initial mass of the meteoroid for asteroidal fireballs in predefined bins. Numbers and colors designate the percentage of the entry mass destroyed at a certain aerodynamic pressure. See Table~\ref{oc_tab} for the details and identification of the asteroidal fireballs.}
      \label{stren_dist_oc}
\end{figure}

The model allows us to derive a proxy for strength distribution inside meteoroids. We used the procedure described in Sect.~\ref{mech_strength} for Geminids and asteroidal fireballs and plotted two types of graphs. The first is a histogram of the distribution of the aerodynamic pressure causing the fragmentation of the meteoroid and its fragments. The distribution is binned in predefined pressure bins (four per order of magnitude) and presented as a heatmap that enables a good comparison of all the meteoroids at a glance. These graphs are shown in Figs.~\ref{stren_dist_gem} and~\ref{stren_dist_oc}.

\begin{figure*}
   \centering
   \includegraphics[width=\hsize]{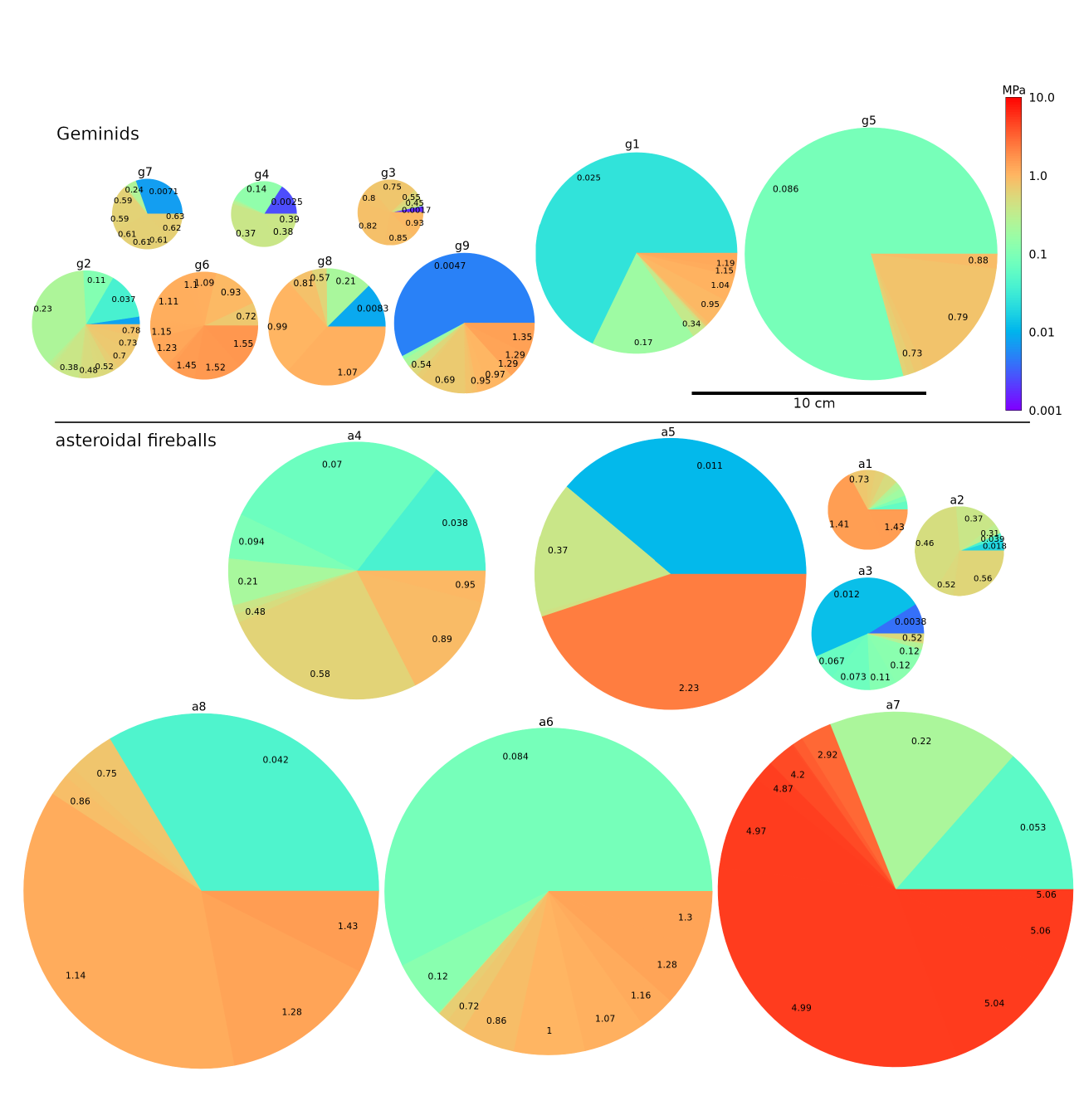}
      \caption{Aerodynamic pressure distribution for Geminids (upper part) and asteroidal fireballs (lower part). Numbers and colors designate the actual aerodynamic pressure of fragmentation in MPa; the fractions represent the destroyed mass. The size of the pie chart is proportional to the size of the meteoroids assuming a spherical shape and $\rho_{\rm bulk}=2500\,\rm{kg\,m^3}$ for Geminids and $\rho_{\rm bulk}=3500\,\rm{kg\,m^3}$ for the asteroidal fireballs. See Table~\ref{gem_tab} for identification of Geminids and Table~\ref{oc_tab} for identification of the asteroidal fireballs.}
      \label{stren_pie}
\end{figure*}

The second type of graph is a pie chart that uses the actual values of aerodynamic pressure (shown in the plot) and the portion of mass destroyed at that pressure. These plots enable a detailed view of individual meteoroids, reveal similarities in strength distribution, and assess the size effects. They are shown in Fig.~\ref{stren_pie} for modeled Geminids and also for the modeled asteroidal fireballs for comparison.

\subsection{Mass loss modes}
\begin{figure}
   \centering
   \includegraphics[width=\hsize]{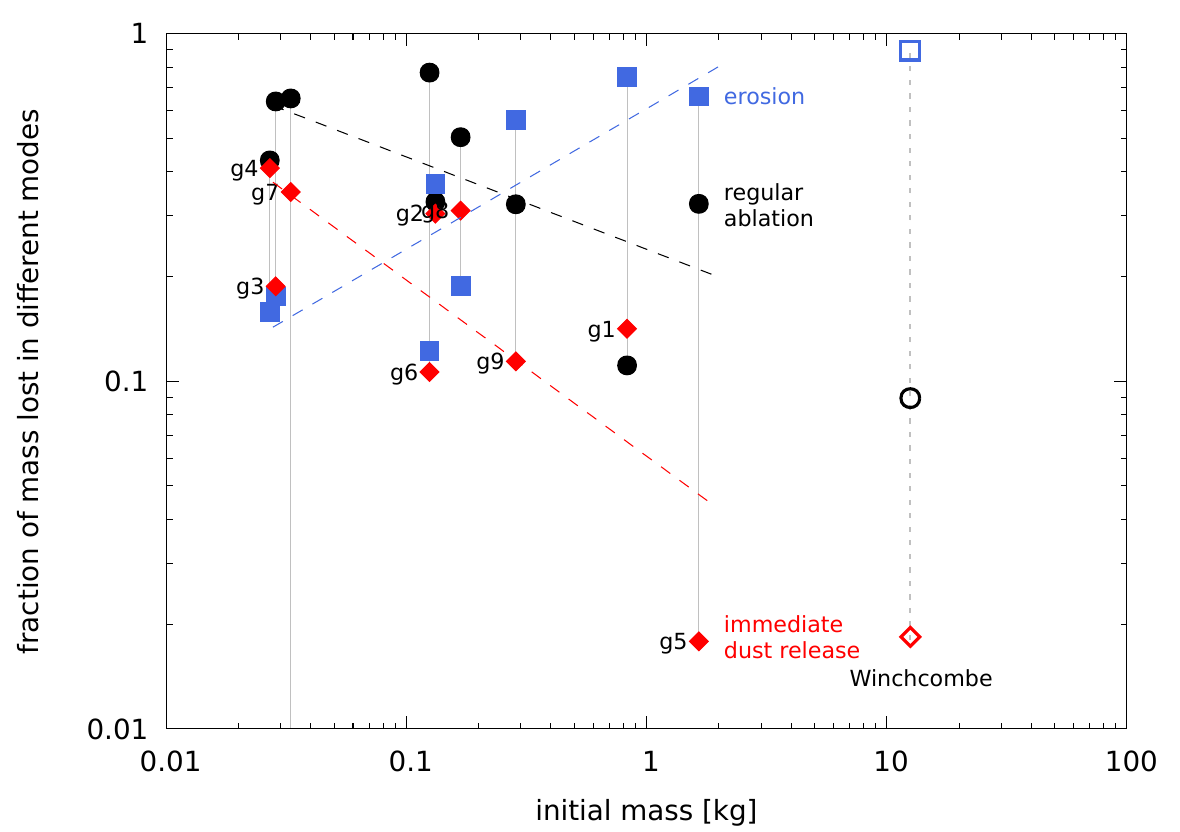}
      \caption{Fraction of mass lost in different modes for Geminids and for the Winchcombe fireball. Both axes are logarithmic. The power-law fits do not include the Winchcombe fireball. See Table~\ref{gem_tab} for Geminid identifications.}
      \label{maslost_gem}
\end{figure}

\begin{figure}
   \centering
   \includegraphics[width=\hsize]{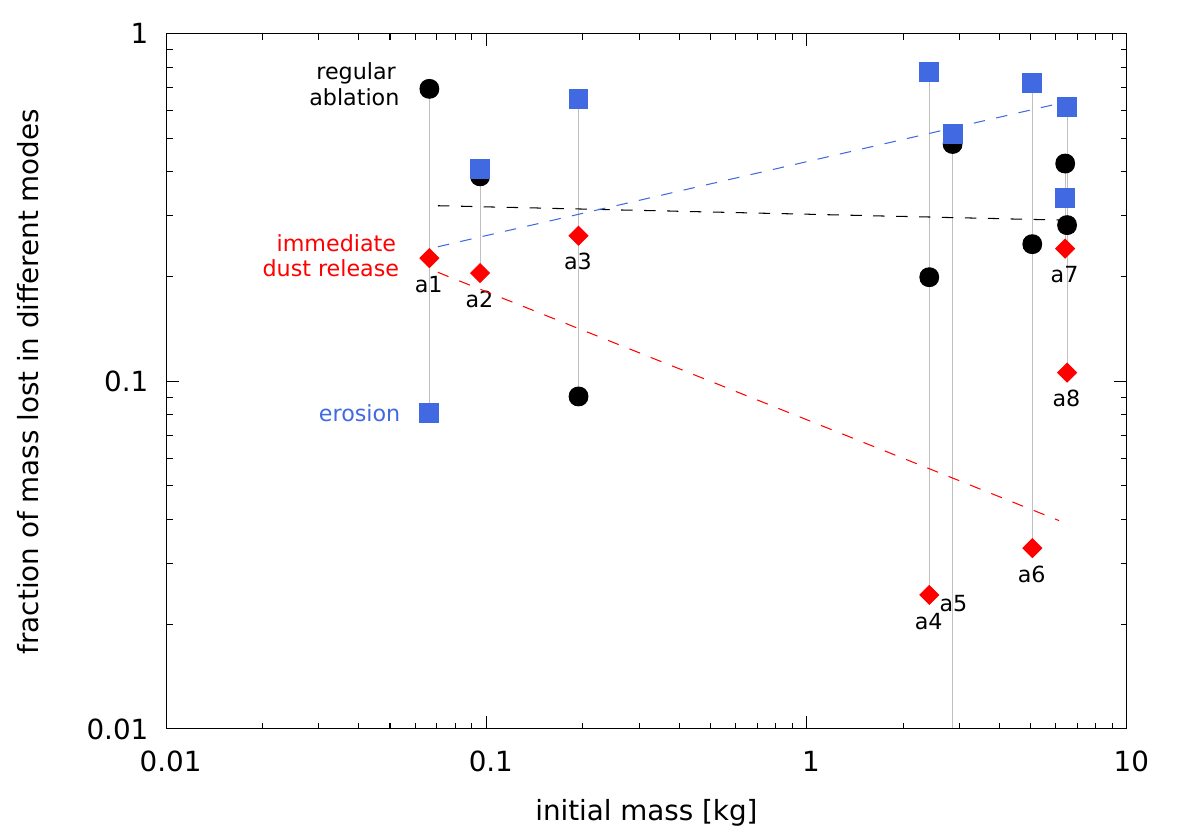}
      \caption{Fraction of mass lost in different modes for asteroidal fireballs. Both axes are logarithmic. See Table~\ref{oc_tab} for identification of the fireballs.}
      \label{maslost_oc}
\end{figure}

Figures~\ref{maslost_gem} and~\ref{maslost_oc} show a fraction of mass lost in different modes versus the initial mass of the meteoroid (see Sect.~\ref{massloss} for details). They also show power-law fits of the trends in these three modes. In Geminids, the amount of mass consumed by an immediate dust release and regular ablation decreases with increasing initial meteoroid mass. Conversely, the importance of erosion of dust grains increases with increasing initial meteoroid mass. While the trends are the same as in the case of Taurids \citep{borovicka2020a}, regular ablation is more important than immediate dust release when compared to Taurids. Asteroidal fireballs show similar trends as Geminids, but they are less pronounced and the fractions vary greatly for individual fireballs.

\subsection{Mass evolution}
\begin{figure}
   \centering
   \includegraphics[width=\hsize]{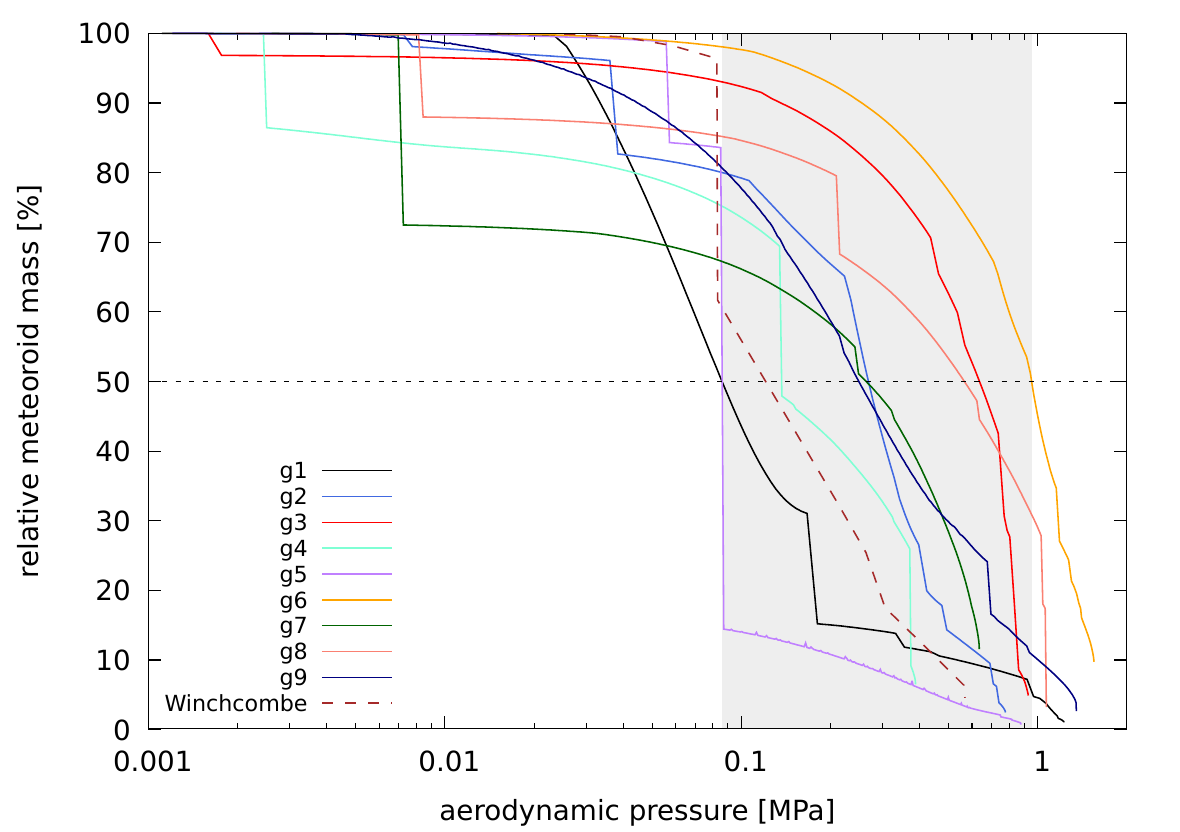}
      \caption{Geminids relative total mass evolution vs. aerodynamic pressure. The shaded area designates the span of the pressures where the meteoroids lost $50\%$ of the entry mass. Also included is the Winchcombe fireball for comparison. See Table~\ref{gem_tab} for Geminid identifications.}
      \label{massevo_gem}
\end{figure}

\begin{figure}
   \centering
   \includegraphics[width=\hsize]{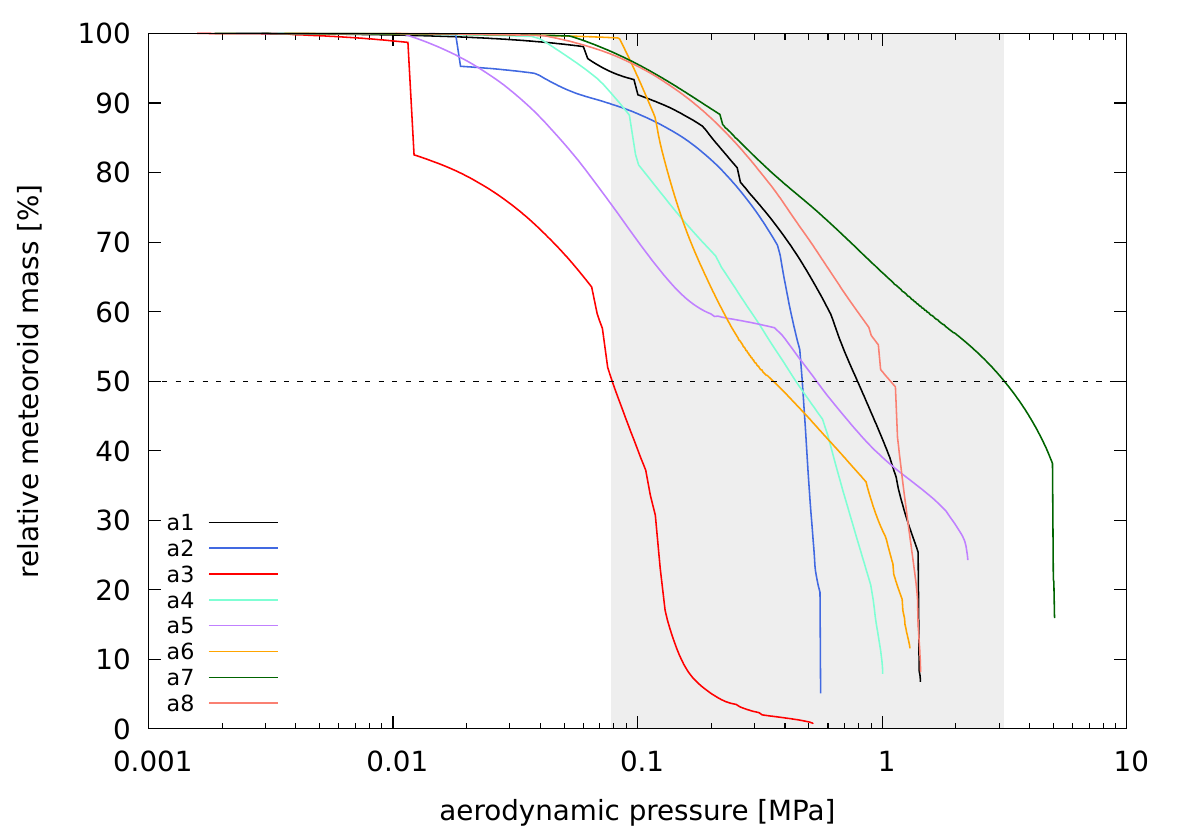}
      \caption{Relative total mass evolution vs. aerodynamic pressure of the asteroidal fireballs. The shaded area designates the span of the pressures where the meteoroids lost $50\%$ of the entry mass. See Table~\ref{oc_tab} for identification of the fireballs.}
      \label{massevo_oc}
\end{figure}

Figures~\ref{massevo_gem} and~\ref{massevo_oc} show the evolution of the total meteoroid mass as a function of aerodynamic pressure that was exerted on its front part calculated by a procedure described in Sect.~\ref{massevo}. It includes the whole mass of the meteoroid at a specific timestep of the model, all the fragments, and dust grains. The aerodynamic pressure on the horizontal axis is the maximum pressure acting on any fragment at the timestep. We also plotted a range of aerodynamic pressures, as a shaded rectangle, at which the meteoroid lost $50\%$ of its entry mass. For Geminids, it is between $0.086\,\rm{MPa}$ and $0.95\,\rm{MPa}$ and for asteroidal fireballs it is between $0.078\,\rm{MPa}$ and $3.15\,\rm{MPa}$.

\subsection{Aerodynamic pressure statistics}
\begin{figure}
   \centering
   \includegraphics[width=\hsize]{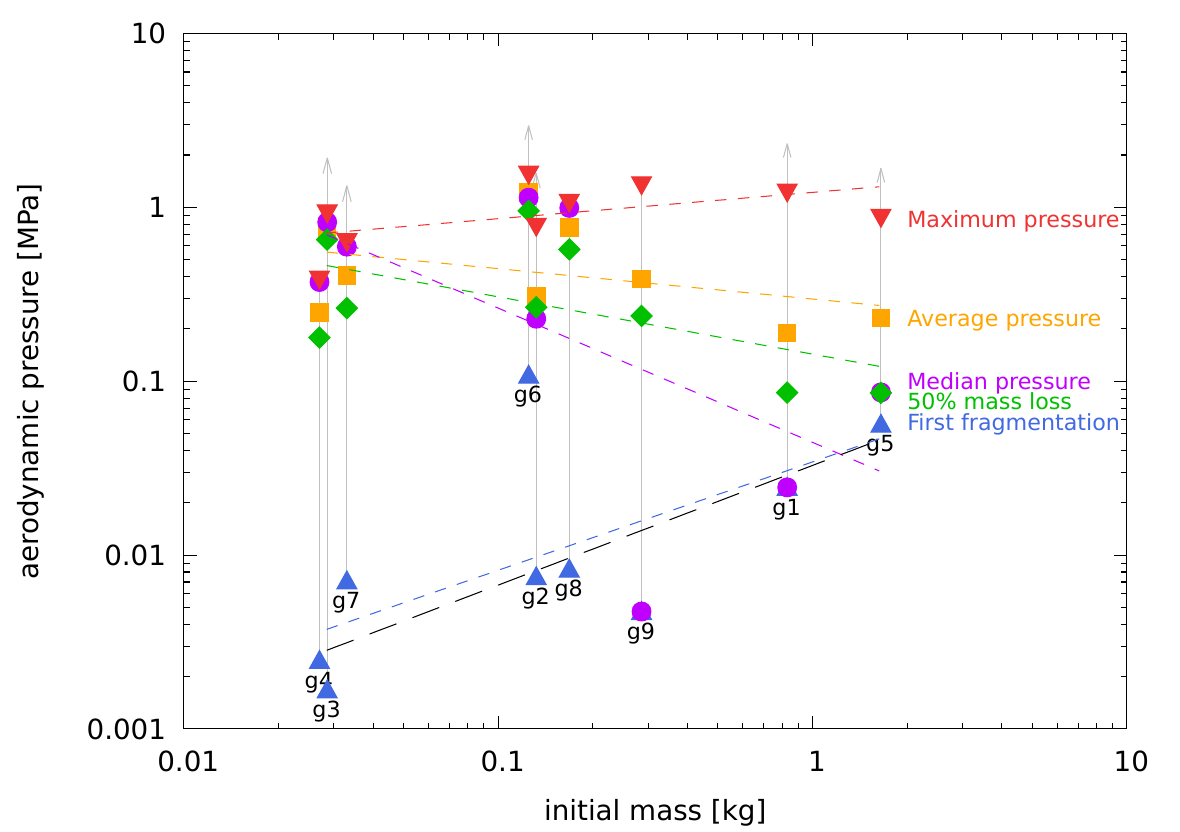}
      \caption{Statistical values of aerodynamic pressure for the modeled Geminids as a function of their initial mass. Both axes are logarithmic. The gray arrows at the maximum pressure indicate a lower limit of the pressure. Also shown are power-law fits of the displayed characteristics. The dashed lines include all the Geminids, while the long-dashed black line at the bottom of the plot excludes Geminid~6. See Table~\ref{gem_tab} for Geminid identifications.}
      \label{dyn_press_stat_gem}
\end{figure}

\begin{figure}
   \centering
   \includegraphics[width=\hsize]{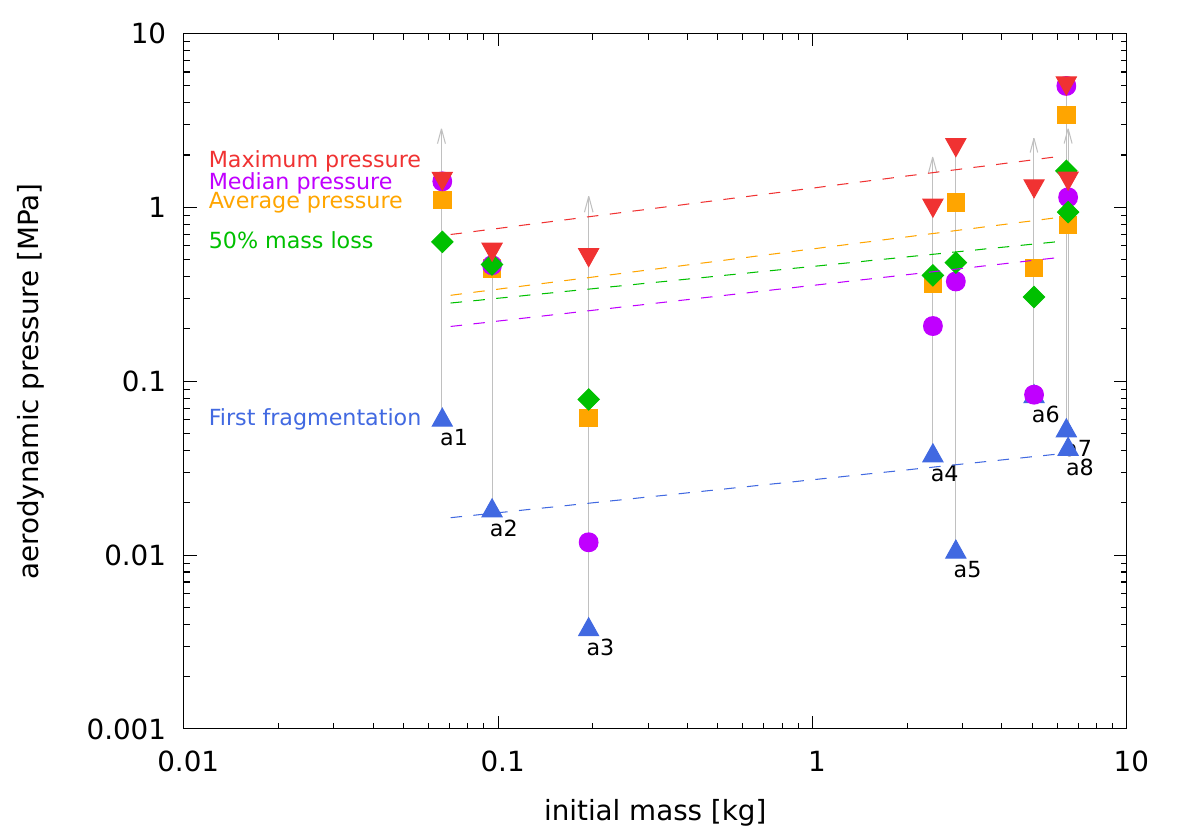}
      \caption{Statistical values of aerodynamic pressure for the modeled asteroidal fireballs as a function of their initial mass. Both axes are logarithmic. The gray arrows at the maximum pressure indicate a lower limit of the pressure. Also shown are power-law fits of the displayed characteristics. See Table~\ref{oc_tab} for identification of the fireballs.}
      \label{dyn_press_stat_oc}
\end{figure}

We can also plot summary statistical values of the aerodynamic pressure acting on each Geminid meteoroid to look for some interesting trends and to obtain an overall picture. In Fig.~\ref{dyn_press_stat_gem} we plot a pressure at the first fragmentation event, an average and median aerodynamic pressures, a pressure at which half of the entry mass is lost, and also a maximum pressure attained by any meteoroid fragment for all modeled Geminids as a function of their initial mass. Figure~\ref{dyn_press_stat_oc} shows the same values for the asteroidal fireballs. The distribution of aerodynamic pressure fragmenting the meteoroid is non-Gaussian rendering the average aerodynamic pressure value less reliable. The gray arrows at the maximum pressure value indicate a lower limit of this value reached by a non-fragmenting part of the meteoroid. The actual pressure that would break such a fragment would be higher.

\subsection{Pressure factor criterion}
The pressure resistance factor (or pressure factor) is defined in \citet{borovicka2022b} as

\begin{equation}
  P\!f=100\,p_{\rm max}\,\cos^{-1}{z}\,m^{-1/3}_{\rm phot}\,v^{-3/2}_{\infty},
\end{equation}where $p_{\rm max}$ is the maximum aerodynamic pressure attained by the meteoroid in MPa, $z$ is the average zenith distance of the radiant, $m_{\rm phot}$ is the initial photometric mass in kg, and $v_{\infty}$ is the entry velocity in $\rm{km\,s^{-1}}$. The pressure factor is roughly proportional to the median mechanical strength of the meteoroid. The traditionally used PE criterion, based on fireball end height \citep{ceplecha1976}, rather depends on the strength of the strongest part of the meteoroid.

\begin{figure}
   \centering
   \includegraphics[width=\hsize]{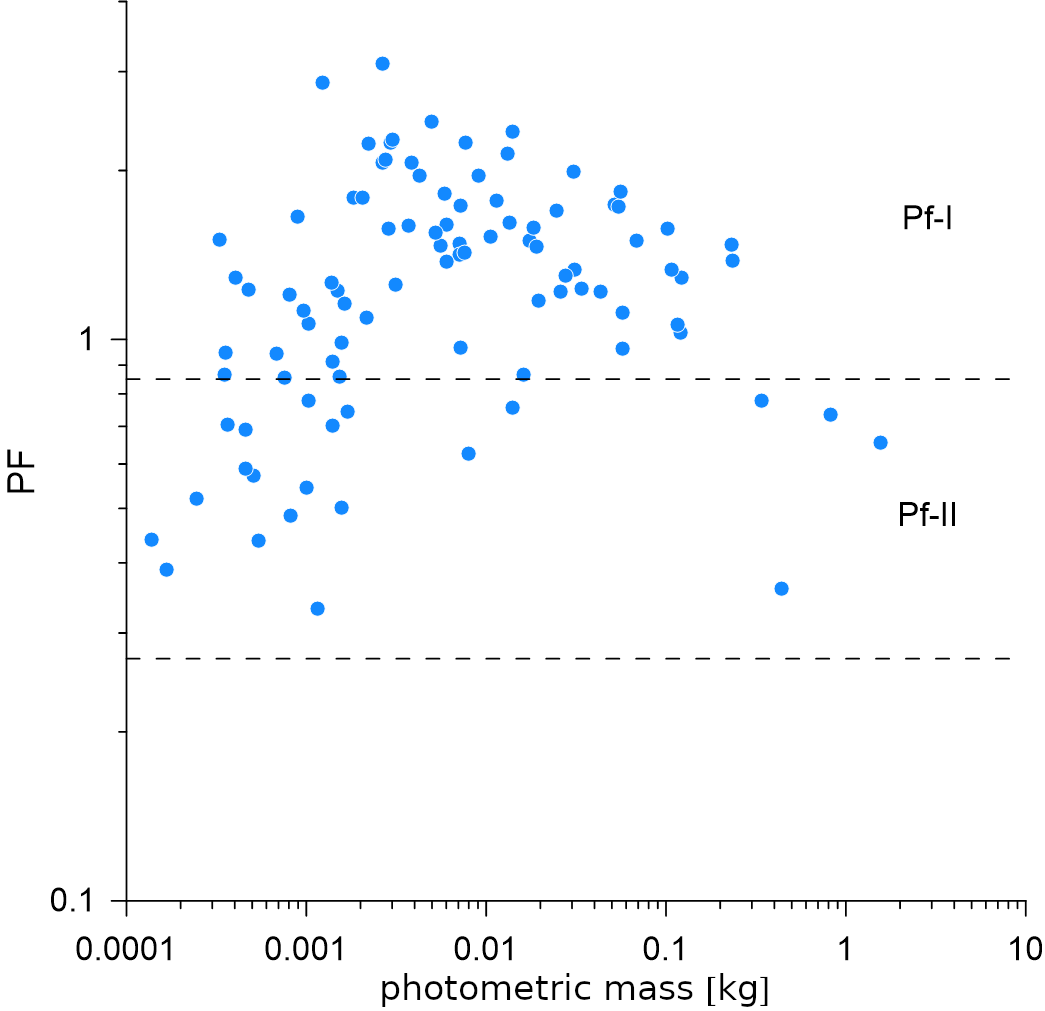}
      \caption{Pressure factor vs. meteoroid photometric mass for Geminid fireballs observed by the EN between 2016 and 2022.}
      \label{gem_pf}
\end{figure}

In Fig.~\ref{gem_pf}, we plot it for a larger sample of Geminid fireballs observed by the EN. The majority of these fireballs were not suitable for detailed modeling. The main reason for this is the lower quality of the radiometric light curve for fainter fireballs, which is crucial for successful modeling, and also the lower amount of dynamical data for short-lived fireballs (or fireballs at a large distance from the EN cameras). It is necessary to carefully select the fireballs to be modeled in order to obtain well-justified results.

However, without detailed modeling, we can at least plot the maximum pressure reached in their trajectory calculated from a dynamical solution corrected for differences in the photometric masses, velocities, and zenith distances of the meteoroids. We note that the photometric mass of a meteoroid is an estimate of its entry mass, which we can derive more accurately with detailed modeling.

\subsection{Spectra of Geminids}
\begin{figure}
   \centering
   \includegraphics[width=\hsize]{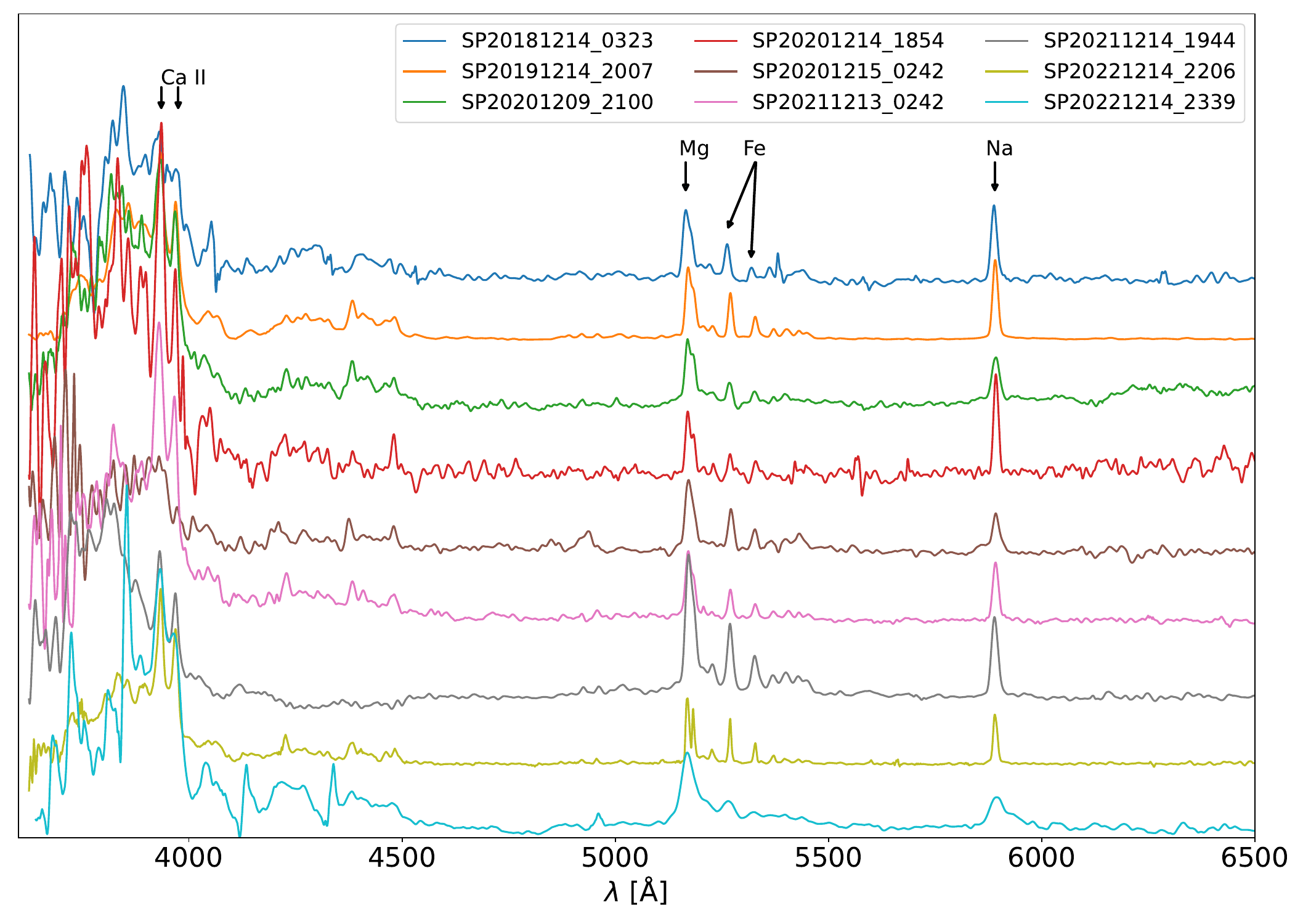}
      \caption{Visible spectra observed by the EN for nine Geminid fireballs, seven of which were modeled in this study. Calcium, magnesium, iron, and sodium lines are indicated. The sodium line is clearly visible in all cases.}
      \label{spectra_gem}
\end{figure}

Nine Geminid fireballs were also observed spectroscopically between 2018 and 2022 with high enough quality to derive intensities of a few spectral lines. Seven of these fireballs were modeled in this study. We observe only minor variability of sodium line intensity in these centimeter-sized Geminid fireballs (Fig.~\ref{spectra_gem}). The analysis of the sodium, magnesium, and iron line intensities suggests normal spectra for all these Geminids.

\section{Discussion}\label{discussion} 
Figs.~\ref{pres_mass_gem} and~\ref{pres_rel_mass_gem} serve as a comparison of aerodynamic pressure causing fragmentation of Geminids to the aerodynamic pressure of fragmentation of asteroidal fireballs and carbonaceous material of the Winchcombe fireball (CM2 meteorites). This comparison of breakup aerodynamic pressures suggests that Geminids might be composed of a more compact and coherent carbonaceous material than CM2. That is also consistent with the spectral composition of their parent body, Phaethon, which is an active B-type asteroid with a negative spectral slope \citep{bus2002,jewitt2010}. For these B-type asteroids, the connection with carbonaceous chondrite meteorites (Pallas group; CV, CO, and CK meteorites) is asserted \citep{clark2010}. The connection of Phaethon with (2)~Pallas was suggested by spectral similarities and by dynamical solutions \citep{deleon2010,todorovic2018}, but it was recently disputed in \citet{maclennan2021}. Instead, \citet{maclennan2022lpi} suggested that Phaethon specifically was related to the Yamoto-group (CY) carbonaceous chondrites, aqueously altered primitive material that was later thermally transformed.

We observe various distributions of aerodynamic pressure causing fragmentation of Geminids in our limited sample. Nevertheless, there are some general characteristics of these meteoroids. First, we look at Figs.~\ref{stren_dist_gem}, \ref{stren_dist_oc}, and~\ref{stren_pie} in Sect.~\ref{results}.

Small Geminids with entry masses below about $0.29\,\rm{kg}$ release some amount (and more than $50\%$ in one case) of dust at very low aerodynamic pressures ($0.001\rm{-}0.01\,\rm{MPa}$), then there is a quiet period without fragmentation (with the ongoing erosion of dust grains) and after that, the majority of the meteoroid mass is destroyed at rather high pressure ($0.5\rm{-}1.5\,\rm{MPa}$).

To the contrary, larger meteoroids with an initial mass higher than about $0.5\,\rm{kg}$ (Geminids~1 and~5) withstand an order of magnitude higher pressures than small meteoroids before they start to crumble, but the large part of the meteoroid is fragmented at this pressure. The rest of the meteoroids reach a maximum pressure similar to that of the small meteoroids (see also Fig.~\ref{dyn_press_stat_gem}). This means that they are mostly composed of a relatively weak material ($0.02\rm{-}0.1\,\rm{MPa}$) with some amount of a rather strong material.

There are two exceptional meteoroids in our sample. Geminid~6, with the initial mass of $0.125\,\rm{kg}$, does not contain any weak part. The first fragmentation occurs at $0.72\,\rm{MPa}$ and the largest pressure causing fragmentation is $1.55\,\rm{MPa}$. The great majority of its mass fragments at the aerodynamic pressure $>1\,\rm{MPa}$. It could have been some more coherent pebble that does not contain too many substantial cracks. Moreover, it breaks the obvious trend of increasing first fragmentation pressure with increasing mass of the meteoroid seen in Fig.~\ref{dyn_press_stat_gem}. We note that this Geminid was observed in poorer weather conditions and from a greater distance, which could have affected the observations of the beginning of the fireball. Geminid~2, with an entry mass of $0.132\,\rm{kg}$, fragments heavily at various aerodynamic pressure values losing a comparable amount of mass at each fragmentation.

The distribution of aerodynamic pressure that fragments asteroidal fireballs varies more than in Geminids because the fireballs are a random sample of sporadic background. Additionally, the fireballs used for the modeling were chosen to be similar to Geminids in terms of entry masses and some also in velocities, masses and velocities that are less typical for asteroidal orbits \citep{borovicka2022b}. Therefore, we could have expected the modeled meteoroids would come from various sources and would have different mechanical properties as well.

There are some similarities between Geminids and asteroidal fireballs, compare for example Geminid~1 and the asteroidal fireball~6 (a6). The distributions look quite similar although the values of breakup pressures are not the same, the first fragmentation of Geminid~1 occurs at an  aerodynamic pressure $3.4$ times lower than in the a6 and it reaches a similar maximum pressure. Their entry mass ratio is~1:6, and by extrapolation of the trend of the minimum aerodynamic pressure for Geminids we obtain a value of $0.1\,\rm{MPa,}$ which is close to the observed value of the first fragmentation for a6 ($0.084\,\rm{MPa}$).

Figures~\ref{maslost_gem} and~\ref{maslost_oc} show what process is mostly responsible for the mass loss in Geminids and asteroidal fireballs of different entry masses. Based on the comparison with the same plot for Taurids \citep{borovicka2020a}, we suggest that this plot can be used in discerning between cometary and asteroidal material for a sample of several fireballs of the same shower. We note that the power-law fits only show trends observed in the sample data and should not be extrapolated. More data are needed to understand this behavior over a larger range of meteoroid initial masses.

The median (average and 50\% mass loss) aerodynamic pressure of Geminids in Fig.~\ref{dyn_press_stat_gem} decreases with increasing entry mass of the meteoroid, which is an expected behavior and also observed in Earth rocks \citep{weibull1939,weibull1951,holsapple2009}.

It is probably more interesting to note that the trend of the minimum aerodynamic pressure at which the Geminids start to crumble is possibly consistent with a theoretical prediction of \cite{capek2011,capek2012lpi} of preferential removal of larger and weaker meteoroids due to thermal stresses. They simulated bodies with a tensile strength of $2\,\rm{MPa}$ (on the order of what we observe), with three different meteoroid sizes of $1\,\rm{mm}$, $1\,\rm{cm}$, and $1\,\rm{dm}$. While the population of millimeter-sized meteoroids was intact, the centimeter-sized population was depleted in weaker members and the decimeter-sized population was even more affected.

The model is described in detail in \cite{capek2010} where a simplified approach is employed and the authors predict preferential destruction of larger and weaker Geminid meteoroids. \cite{capek2012} developed a more sophisticated model that deals with a creation of an insulating layer of material formed by thermal stresses that prevent further destruction of meteoroids. This less coherent material would be destroyed by much smaller aerodynamic pressure than the intact core of the meteoroid during the atmospheric passage. This is perhaps what we observe for the two largest Geminids, but our sample is too limited to allow any firm conclusions to be drawn.

When we compare the median aerodynamic pressure causing fragmentation of the fireballs we modeled in Fig.~\ref{dyn_press_stat_gem} (entry masses between $0.027\,\rm{kg}$ and $1.7\,\rm{kg}$) and the pressure factor for these fireballs (Fig.~\ref{gem_pf}), we observe the same trend. This is because the pressure factor is a global measure characterizing the overall strength of the meteoroid. For meteoroid masses larger than about $2\,\rm{g}$, the pressure factor decreases with increasing mass, so the larger Geminids are overall softer though they contain strong parts. For the mass of $2\,\rm{g}$, the pressure factor is maximum, and then it drops very fast for smaller masses. If mechanical forces destroyed the small Geminids, the pressure factor plot suggests they were as weak as the largest Geminids observed by the EN. But more research is needed to describe the transition between mechanical and thermal destruction of small meteoroids.

We note that our conclusions are based on a very limited sample of nine Geminids. We observe trends in the data that we believe are generally valid, but they may be refined as more high-quality data on Geminid fireballs become available. Therefore, all the conclusions we make are preliminary.

Also, the model fits are not perfect. Figure~\ref{gem6_rlc} shows a radiometric light curve of the EN131218 together with the physical model. In the time interval of $t=-0.1\rm{-}0.3\,\rm{s}$ the lower slope of the model is caused by the fact that the meteoroid is probably in the preheating phase and just starting to shine. Our model does not include a comprehensive physical description of this phase.

Other deviations of the fit can be noted at times $\sim0.6\,\rm{s}$ and $\sim0.9\,\rm{s}$. It seems that adding more eroding grains could fix the discrepancy, but that would require more mass, which could lead to another discrepancy in the earlier part of the light curve. Nevertheless, even if the model is not perfect, the most important parameters, namely fragmentation heights and fragment masses, are robust enough.

We were interested how our values of a proxy for mechanical strength compare to previously derived strengths of Geminids. \citet{trigo2006} attempted to derive this quantity for meteoroids related to major meteor showers, including Geminids. They obtained an average value of $2.2\pm0.2\cdot10^4\,\rm{Pa}$, which is close to the lower limit of our aerodynamic pressure values when larger Geminids only start to fragment. However, the fragmentation of small meteoroids is likely not caused by mechanical forces as \citet{trigo2006} assumed, but rather by thermal stresses. Moreover, the authors related the point of maximum observed brightness to the fragmentation point and this is probably not the case \citep{borovicka2007}. The analysis they used is only useful for larger meteoroids where we can observe gross fragmentation events marked by a bright flare in their light curve.

\citet{madiedo2013} observed and analyzed a bright Geminid fireball with a photometric mass of $\sim\!0.76\,\rm{kg}$ and from the observed light curve derived the tensile strength proxy at the fragmentation event related to the brightest flare of $3.8\pm0.4\,\rm{MPa}$. This value is higher than any value of the maximum aerodynamic pressure reached by Geminid meteoroids with comparable initial mass derived in this study but it is of the same order. For bolides we modeled, the maximum pressure is not attained at the brightest point, but rather in the last third of the event duration (and usually even later, at the last bright fragmentation event) before the surviving meteoroid fragment decelerates substantially. Were it also the case of the bolide analyzed by \citet{madiedo2013}, the maximum aerodynamic pressure would be even higher, making it as strong as common asteroidal meteoroids producing ordinary chondrites.

Another bright Geminid fireball was described in \citet{prieto2013}. They derived the aerodynamic pressure for the two brightest flares observed in the fireball light curve as $2.0\pm0.4\cdot10^4\,\rm{Pa}$ and $3.4\pm0.4\cdot10^4\,\rm{Pa}$. Unfortunately, the authors did not publish the initial mass of the meteoroid, only the maximum brightness of $-10\pm1\,\rm{mag}$ and the initial velocity of $39.0\pm0.3\,\rm{km\,s^{-1}}$. This velocity seems too high for Geminids so we suspect there was some trouble in the calculation. The calculated pressures for this Geminid are too low for the main fragmentation events.

\citet{beech2002} analyzed light curves of three Geminids focusing on high-frequency flickering. He assumed that the flickering was caused by a fast-spinning meteoroid that led to a rotational modulation of ablation. The putative fast rotation caused the Geminids to be destroyed by rotational bursting rather than by aerodynamic pressure. The derived tensile strengths were between $2.3\cdot10^5\,\rm{Pa}$ and $3.2\cdot10^5\,\rm{Pa}$ for a bulk density of $1000\,\rm{kg\,m^{-3}}$. When we use our value of $2500\,\rm{kg\,m^{-3}}$, the strengths are $2.5$ times higher, or $5.8\rm{-}8.0\cdot10^5\,\rm{Pa}$. 

\citet{beech2002} estimated the initial mass of one of the Geminids based upon the maximum observed brightness and theoretical calculations as $0.025\,\rm{kg}$, the initial masses of the other two Geminids are $0.012\,\rm{kg}$ and $0.010\,\rm{kg}$~\citep{baba2004}. The masses of these Geminids are lower than the masses of those we modeled, but the tensile strengths derived by \citet{beech2002} are comparable to the aerodynamic pressures we derived assuming a different fragmentation process. We also observe flickering in Geminid radiometric light curves (see Fig.~\ref{gem6_rlc}) with maximum frequencies on the order of several hundred hertz and the flickering is observed both before and after gross fragmentation events. The radiometric light curves acquired by the EN have a sampling of $5000\,\rm{Hz}$.

Flickering was observed by the EN in many other high-time-resolution radiometric light curves of asteroidal fireballs and even those that produced meteorites. For example, \citet{spurny2012} analyzed the Bunburra Rockhole meteorite fall in Australia observed by the Desert Fireball Network and concluded that the flickering was probably not caused by the rotation of the meteoroid.

\citet{otto2023} calculated the tensile strength from the thermal conductivity measured on the surface of (162173)~Ryugu by the MASCOT lander and in orbit by the thermal mapper of the Hayabusa2 mission. In the calculations, they assumed Young's modulus representative of carbonaceous chondrites. The values derived for a typical observed boulder diameter of $\sim\!\!10\,\rm{cm}$ were $200\rm{-}280\,\rm{kPa}$ and $\sim\!229\,\rm{kPa}$, respectively. These values are in agreement with the values of the tensile strength proxy we derived for Geminids.

\citet{kurosawa2022} measured the flexural strength of a small ($3\,\rm{mm}$) sample of the C-type asteroid (162173)~Ryugu brought to Earth by the Japan Aerospace Exploration Agency (JAXA) Hayabusa2 mission. The spectra of the Ryugu samples and Ryugu itself are close to the CI chondrites \citep{yada2022}. We note that these samples traveled through the atmosphere of Earth in a hermetically sealed container inside the re-entry capsule. Flexural strength is a composite of compressive and tensile strength, but according to \citet{kurosawa2022}, the tested sample was clearly fractured by tensile stress. They found the value of $3\rm{-}8\,\rm{MPa,}$ which is higher by a factor of a few than the aerodynamic pressure that caused the last observed fragmentation of Geminid meteoroids we derived. We recall that strength scales with size and that the smaller samples are stronger than the larger ones.

We can also compare the derived strength proxy of meteoroid fragmentation to the strength measured for carbonaceous chondrite meteorites. Tensile strengths of two such meteorites are given in Table~III of \citet{svetsov1995}, $28\,\rm{MPa}$ for Allende (CV3) and $31\,\rm{MPa}$ for Kainsaz (CO3). \citet{tsuchiyama2008} measured tensile strengths of $100\,\rm{\mu m}$ samples of several carbonaceous chondrites, namely Murchison (CM2, $2.0\pm1.5\,\rm{MPa}$), Ivuna (CI1, $0.7\pm0.2\,\rm{MPa}$), Orgueil (CI1, $2.8\pm1.9\,\rm{MPa}$), Tagish Lake (carbonate-poor sample, C2-ung, $0.8\pm0.3\,\rm{MPa}$), Tagish Lake (carbonate-rich sample, C2-ung, $6.7\pm9.8\,\rm{MPa}$), and Murray (CM2, $8.8\pm4.8\,\rm{MPa}$). The maximum strength proxy values of Geminids we derived by the atmospheric fragmentation modeling are $0.4\rm{-}1.6\,\rm{MPa}$. These values are comparable to the weakest measured meteorite grains and a factor of a few weaker than the strongest grains. That also supports our main conclusion that Geminids are made of a more coherent carbonaceous material.

Spectra of all observed Geminid fireballs are classified as normal with little variation of sodium content. To the contrary, spectra of small Geminid meteors are known for high variability in the abundance of sodium \citep{borovicka2005,vojacek2015,abe2020} oscillating between the normal classification of the spectrum to sodium-free spectra. This observation supports the idea of the thermal removal of sodium during perihelion passages \citep{capek2009}. For monolithic meteoroids, this process is much faster for small millimeter-sized bodies observed by video cameras and slower for larger centimeter-sized meteoroids. For porous meteoroids the dependence is more complicated, with lower sodium content expected for more porous Geminids with smaller grains. For small Geminids we observe varying porosity, while larger meteoroids tend to have lower porosity. Therefore, sodium-free Geminid fireballs are less likely to be observed than fainter Geminid meteors.

\section{Conclusions}\label{conclusions} 
We calculated a detailed fragmentation model of nine Geminid fireballs in the mass range $0.027\rm{-}1.7\,\rm{kg}$ and three asteroidal fireballs, which were supplemented with another five from a previous study \citep{henych2023} in the mass range $0.066\rm{-}6.5\,\rm{kg}$ for comparison. The modeling enabled us to derive the aerodynamic pressure at which the meteoroids fragment and therefore place constraints on their mechanical strength and physical character. We find that Geminids can withstand higher aerodynamic pressure in the final stages of fragmentation than the Winchcombe meteoroid that produced carbonaceous chondrites but lower pressures than asteroidal fireballs of similar entry masses. Moreover, Phaethon, the parent body of the Geminids, is a B-type asteroid, part of the C complex. This complex is characterized by a primitive carbonaceous composition that is probably related to some types of carbonaceous chondrites. Therefore, we conclude that the Geminid meteoroids are composed of a compact and coherent carbonaceous material that is stronger than CM2 material.

We also derived the distribution of aerodynamic pressures that cause the fragmentation of Geminids and asteroidal fireballs. We presume that the strength distribution inside the meteoroids is proportional to the aerodynamic pressures. Small Geminids ($m\lesssim0.29\,\rm{kg}$) start to crumble at very low pressures of $0.001\rm{-}0.01\,\rm{MPa}$, but the majority of their mass is destroyed at the rather high pressures of $0.5\rm{-}1.5\,\rm{MPa}$. The larger Geminids ($m\gtrsim0.5\,\rm{kg}$) are mostly composed of a relatively weak material that fragments at $0.02\rm{-}0.1\,\rm{MPa}$ but also contain some stronger material whose strength proxy was comparable to that of the small Geminids. For some asteroidal fireballs, we derived an aerodynamic pressure distribution that is similar to that of the Geminids, suggesting that these asteroidal fireballs may also have been composed of a carbonaceous material.

The mass loss regimes in Geminids show similar trends as in Taurids. The importance of erosion increases with increasing entry mass of the meteoroid, and both regular ablation and immediate dust release become less important for larger meteoroids. Regular ablation is a more important mass loss mode than immediate dust release for Geminids, but the opposite is true for Taurids. This could help us draw a line between asteroidal (Geminid) and cometary (Taurid) material.

The minimum aerodynamic pressure increases with increasing initial mass of a Geminid meteoroid. To the contrary, the median (average, $50\%$ mass loss) aerodynamic pressure that fragments Geminids decreases with increasing entry mass, which is consistent with the simplified analysis of the overall mechanical strength via the pressure factor (see Fig.~\ref{gem_pf}) for Geminids with photometric masses $\gtrsim2\,\rm{g}$.

Geminid physical properties reflect not only the material properties of Phaethon but also their subsequent evolution after being released from Phaethon (by a still unknown mechanism). Because of their low perihelion distance, Geminid meteoroids are subject to thermal stresses, which are more important for larger bodies \citep{capek2010,capek2012}. The observed meteoroids with sizes of a few centimeters have a bimodal structure with a minor weak part and a larger strong part. This suggests that Phaethon is a mixture of granular material and a compact carbonaceous material. Meteoroids larger than about $5\,\rm{cm}$ do not contain the granular material. They were probably subject to additional fragmentation in space caused by thermal stresses, leading to the loss of the surface layers that would have contained the granular material. Moreover, thermal stresses fractured most of the remaining compact material, diminishing the overall bulk strength of larger meteorites. \citet{borovicka2010} studied millimeter-sized meteoroids and found they were mostly composed of granular material, with a fundamental grain size of $80\rm{-}300\,\rm{\mu m}$.

For nine Geminids, high-quality spectra were obtained. They show much lower variations in sodium content than was observed for smaller Geminid meteoroids. This conforms with the prediction of numerical models focused on sodium release from meteoroids orbiting close to the Sun \citep{capek2009}. The other spectral lines observed in the spectra suggest normal spectra for all observed Geminid fireballs.

\begin{acknowledgements}
We thank the anonymous referee for his/her questions and comments that helped us improve the original manuscript. This research was supported by grant no. 19-26232X from the Czech Science Foundation. The computations were performed on the OASA and VIRGO clusters of the Astronomical Institute of the Czech Academy of Sciences. This research has made use of NASA’s Astrophysics Data System.
\end{acknowledgements}

\bibliographystyle{aa}
\bibliography{geminids}

\begin{thebibliography}{88}
\expandafter\ifx\csname natexlab\endcsname\relax\def\natexlab#1{#1}\fi

\bibitem[{{Abe} {et~al.}(2020){Abe}, {Ogawa}, {Maeda}, \& {Arai}}]{abe2020}
{Abe}, S., {Ogawa}, T., {Maeda}, K., \& {Arai}, T. 2020, \planss, 194, 105040

\bibitem[{{Babadzhanov}(2002)}]{baba2002}
{Babadzhanov}, P.~B. 2002, \aap, 384, 317

\bibitem[{{Babadzhanov} \& {Kokhirova}(2009)}]{baba2009}
{Babadzhanov}, P.~B. \& {Kokhirova}, G.~I. 2009, \aap, 495, 353

\bibitem[{{Babadzhanov} \& {Konovalova}(2004)}]{baba2004}
{Babadzhanov}, P.~B. \& {Konovalova}, N.~A. 2004, \aap, 428, 241

\bibitem[{{Beech}(2002)}]{beech2002}
{Beech}, M. 2002, \mnras, 336, 559

\bibitem[{{Beech} {et~al.}(2003){Beech}, {Illingworth}, \&
  {Murray}}]{beech2003}
{Beech}, M., {Illingworth}, A., \& {Murray}, I.~S. 2003, \maps, 38, 1045

\bibitem[{{Belkovich} \& {Ryabova}(1989)}]{belkovich1989}
{Belkovich}, O.~I. \& {Ryabova}, G.~O. 1989, Solar System Research, 23, 98

\bibitem[{{Borovi{\v{c}}ka} {et~al.}(2005){Borovi{\v{c}}ka}, {Koten},
  {Spurn{\'y}}, {Bo{\v{c}}ek}, \& {{\v{S}}tork}}]{borovicka2005}
{Borovi{\v{c}}ka}, J., {Koten}, P., {Spurn{\'y}}, P., {Bo{\v{c}}ek}, J., \&
  {{\v{S}}tork}, R. 2005, \icarus, 174, 15

\bibitem[{{Borovi{\v{c}}ka} {et~al.}(2010){Borovi{\v{c}}ka}, {Koten},
  {Spurn{\'y}}, {{\v{C}}apek}, {Shrben{\'y}}, \& {{\v{S}}tork}}]{borovicka2010}
{Borovi{\v{c}}ka}, J., {Koten}, P., {Spurn{\'y}}, P., {et~al.} 2010, in Icy
  Bodies of the Solar System, ed. J.~A. {Fernandez}, D.~{Lazzaro},
  D.~{Prialnik}, \& R.~{Schulz}, Vol. 263, 218--222

\bibitem[{{Borovi{\v{c}}ka} \& {Spurn{\'y}}(2020)}]{borovicka2020a}
{Borovi{\v{c}}ka}, J. \& {Spurn{\'y}}, P. 2020, \planss, 182, 104849

\bibitem[{{Borovi{\v{c}}ka} {et~al.}(2007){Borovi{\v{c}}ka}, {Spurn{\'y}}, \&
  {Koten}}]{borovicka2007}
{Borovi{\v{c}}ka}, J., {Spurn{\'y}}, P., \& {Koten}, P. 2007, \aap, 473, 661

\bibitem[{{Borovi{\v{c}}ka} {et~al.}(2020){Borovi{\v{c}}ka}, {Spurn{\'y}}, \&
  {Shrben{\'y}}}]{borovicka2020b}
{Borovi{\v{c}}ka}, J., {Spurn{\'y}}, P., \& {Shrben{\'y}}, L. 2020, \aj, 160,
  42

\bibitem[{{Borovi{\v{c}}ka} {et~al.}(2022{\natexlab{a}}){Borovi{\v{c}}ka},
  {Spurn{\'y}}, \& {Shrben{\'y}}}]{borovicka2022b}
{Borovi{\v{c}}ka}, J., {Spurn{\'y}}, P., \& {Shrben{\'y}}, L.
  2022{\natexlab{a}}, \aap, 667, A158

\bibitem[{{Borovi{\v{c}}ka} {et~al.}(2022{\natexlab{b}}){Borovi{\v{c}}ka},
  {Spurn{\'y}}, {Shrben{\'y}}, {{\v{S}}tork}, {Kotkov{\'a}}, {Fuchs},
  {Kecl{\'\i}kov{\'a}}, {Zichov{\'a}}, {M{\'a}nek}, {V{\'a}chov{\'a}},
  {Macourkov{\'a}}, {Svore{\v{n}}}, \& {Mucke}}]{borovicka2022a}
{Borovi{\v{c}}ka}, J., {Spurn{\'y}}, P., {Shrben{\'y}}, L., {et~al.}
  2022{\natexlab{b}}, \aap, 667, A157

\bibitem[{{Boslough} \& {Crawford}(2008)}]{boslough2008}
{Boslough}, M.~B.~E. \& {Crawford}, D.~A. 2008, International Journal of Impact
  Engineering, 35, 1441

\bibitem[{{Bronshten}(1983)}]{bronshten1983}
{Bronshten}, V.~A. 1983, {Physics of Meteoric Phenomena} (Dordrecht, Holland:
  Springer Dordrecht)

\bibitem[{{Bus} \& {Binzel}(2002)}]{bus2002}
{Bus}, S.~J. \& {Binzel}, R.~P. 2002, \icarus, 158, 146

\bibitem[{{Ceplecha} {et~al.}(1998){Ceplecha}, {Borovi{\v{c}}ka}, {Elford},
  {Revelle}, {Hawkes}, {Porub{\v{c}}an}, \& {{\v{S}}imek}}]{ceplecha1998}
{Ceplecha}, Z., {Borovi{\v{c}}ka}, J., {Elford}, W.~G., {et~al.} 1998, \ssr,
  84, 327

\bibitem[{{Ceplecha} \& {McCrosky}(1976)}]{ceplecha1976}
{Ceplecha}, Z. \& {McCrosky}, R.~E. 1976, \jgr, 81, 6257

\bibitem[{{Ceplecha} \& {McCrosky}(1992)}]{ceplecha1992}
{Ceplecha}, Z. \& {McCrosky}, R.~E. 1992, in Asteroids, Comets, Meteors 1991,
  ed. A.~W. {Harris} \& E.~{Bowell}, 109

\bibitem[{{Clark} {et~al.}(2010){Clark}, {Ziffer}, {Nesvorny}, {Campins},
  {Rivkin}, {Hiroi}, {Barucci}, {Fulchignoni}, {Binzel}, {Fornasier}, {DeMeo},
  {Ockert-Bell}, {Licandro}, \& {Moth{\'e}-Diniz}}]{clark2010}
{Clark}, B.~E., {Ziffer}, J., {Nesvorny}, D., {et~al.} 2010, Journal of
  Geophysical Research (Planets), 115, E06005

\bibitem[{{Cukier} \& {Szalay}(2023)}]{cukier2023}
{Cukier}, W.~Z. \& {Szalay}, J.~R. 2023, \psj, 4, 109

\bibitem[{{Davies} {et~al.}(1984){Davies}, {Green}, {Stewart}, {Meadows}, \&
  {Aumann}}]{davies1984}
{Davies}, J.~K., {Green}, S.~F., {Stewart}, B.~C., {Meadows}, A.~J., \&
  {Aumann}, H.~H. 1984, \nat, 309, 315

\bibitem[{{de Le{\'o}n} {et~al.}(2010){de Le{\'o}n}, {Campins}, {Tsiganis},
  {Morbidelli}, \& {Licandro}}]{deleon2010}
{de Le{\'o}n}, J., {Campins}, H., {Tsiganis}, K., {Morbidelli}, A., \&
  {Licandro}, J. 2010, \aap, 513, A26

\bibitem[{{Fadeenko}(1967)}]{fadeenko1967}
{Fadeenko}, Y.~I. 1967, Fizika Goreniya i Vzryva, 3, 276

\bibitem[{{Fox} {et~al.}(1984){Fox}, {Williams}, \& {Hughes}}]{fox1984}
{Fox}, K., {Williams}, I.~P., \& {Hughes}, D.~W. 1984, \mnras, 208, 11P

\bibitem[{{Fujiwara} {et~al.}(1989){Fujiwara}, {Cerroni}, {Davis}, {Ryan}, {di
  Martino}, {Holsapple}, \& {Housen}}]{fujiwara1989}
{Fujiwara}, A., {Cerroni}, P., {Davis}, D.~R., {et~al.} 1989, in Asteroids II,
  ed. R.~P. {Binzel}, T.~{Gehrels}, \& M.~S. {Matthews}, 240--265

\bibitem[{{Gustafson}(1989)}]{gustafson1989}
{Gustafson}, B.~A.~S. 1989, \aap, 225, 533

\bibitem[{{Hanu{\v{s}}} {et~al.}(2018){Hanu{\v{s}}}, {Vokrouhlick{\'y}},
  {Delbo'}, {Farnocchia}, {Polishook}, {Pravec}, {Hornoch},
  {Ku{\v{c}}{\'a}kov{\'a}}, {Ku{\v{s}}nir{\'a}k}, {Stephens}, \&
  {Warner}}]{hanus2018}
{Hanu{\v{s}}}, J., {Vokrouhlick{\'y}}, D., {Delbo'}, M., {et~al.} 2018, \aap,
  620, L8

\bibitem[{{Henych} {et~al.}(2023){Henych}, {Borovi{\v{c}}ka}, \&
  {Spurn{\'y}}}]{henych2023}
{Henych}, T., {Borovi{\v{c}}ka}, J., \& {Spurn{\'y}}, P. 2023, \aap, 671, A23

\bibitem[{{Holsapple}(2009)}]{holsapple2009}
{Holsapple}, K.~A. 2009, \planss, 57, 127

\bibitem[{{Hughes}(1983)}]{hughes1983}
{Hughes}, D.~W. 1983, \nat, 306, 116

\bibitem[{{Hui}(2023)}]{hui2023}
{Hui}, M.-T. 2023, \aj, 165, 94

\bibitem[{{Hui} \& {Li}(2017)}]{hui2017}
{Hui}, M.-T. \& {Li}, J. 2017, \aj, 153, 23

\bibitem[{{Hunt} {et~al.}(1986){Hunt}, {Fox}, \& {Williams}}]{hunt1986}
{Hunt}, J., {Fox}, K., \& {Williams}, I.~P. 1986, in Asteroids, Comets, Meteors
  II, ed. C.~I. {Lagerkvist}, H.~{Rickman}, B.~A. {Lindblad}, \&
  H.~{Lundstedt}, 549

\bibitem[{{Jewitt} \& {Hsieh}(2006)}]{jewitt2006}
{Jewitt}, D. \& {Hsieh}, H. 2006, \aj, 132, 1624

\bibitem[{{Jewitt} \& {Li}(2010)}]{jewitt2010}
{Jewitt}, D. \& {Li}, J. 2010, \aj, 140, 1519

\bibitem[{{Jewitt} {et~al.}(2013){Jewitt}, {Li}, \& {Agarwal}}]{jewitt2013}
{Jewitt}, D., {Li}, J., \& {Agarwal}, J. 2013, \apjl, 771, L36

\bibitem[{{Jones} \& {Hawkes}(1986)}]{jones1986}
{Jones}, J. \& {Hawkes}, R.~L. 1986, \mnras, 223, 479

\bibitem[{{Jones} {et~al.}(2016){Jones}, {Poole}, \& {Webster}}]{jones2016}
{Jones}, J., {Poole}, L.~M.~G., \& {Webster}, A.~R. 2016, \mnras, 455, 3424

\bibitem[{{Kasuga}(2009)}]{kasuga2009}
{Kasuga}, T. 2009, Earth Moon and Planets, 105, 321

\bibitem[{{Kasuga} \& {Jewitt}(2008)}]{kasuga2008}
{Kasuga}, T. \& {Jewitt}, D. 2008, \aj, 136, 881

\bibitem[{{Kasuga} \& {Jewitt}(2019)}]{kasuga2019}
{Kasuga}, T. \& {Jewitt}, D. 2019, in Meteoroids: Sources of Meteors on Earth
  and Beyond, ed. G.~O. {Ryabova}, D.~J. {Asher}, \& M.~J. {Campbell-Brown},
  187

\bibitem[{{Kasuga} \& {Masiero}(2022)}]{kasuga2022}
{Kasuga}, T. \& {Masiero}, J.~R. 2022, \aj, 164, 193

\bibitem[{{Kurosawa} {et~al.}(2022){Kurosawa}, {Tanaka}, {Ino}, {Nakashima},
  {Nakamura}, {Morita}, {Kikuiri}, {Amano}, {Kagawa}, {Yurimoto}, {Noguchi},
  {Okazaki}, {Yabuta}, {Naraoka}, {Sakamoto}, {Tachibana}, {Watanabe}, \&
  {Tsuda}}]{kurosawa2022}
{Kurosawa}, K., {Tanaka}, S., {Ino}, Y., {et~al.} 2022, in LPI Contributions,
  Vol. 2678, 53rd Lunar and Planetary Science Conference, 1378

\bibitem[{{Li} \& {Jewitt}(2013)}]{li2013}
{Li}, J. \& {Jewitt}, D. 2013, \aj, 145, 154

\bibitem[{{Licandro} {et~al.}(2007){Licandro}, {Campins}, {Moth{\'e}-Diniz},
  {Pinilla-Alonso}, \& {de Le{\'o}n}}]{licandro2007}
{Licandro}, J., {Campins}, H., {Moth{\'e}-Diniz}, T., {Pinilla-Alonso}, N., \&
  {de Le{\'o}n}, J. 2007, \aap, 461, 751

\bibitem[{{MacLennan} {et~al.}(2021){MacLennan}, {Toliou}, \&
  {Granvik}}]{maclennan2021}
{MacLennan}, E., {Toliou}, A., \& {Granvik}, M. 2021, \icarus, 366, 114535

\bibitem[{{MacLennan} \& {Granvik}(2022)}]{maclennan2022lpi}
{MacLennan}, E.~M. \& {Granvik}, M. 2022, in LPI Contributions, Vol. 2695, LPI
  Contributions, 6401

\bibitem[{{Madiedo} {et~al.}(2013){Madiedo}, {Trigo-Rodr{\'\i}guez},
  {Castro-Tirado}, {Ortiz}, \& {Cabrera-Ca{\~n}o}}]{madiedo2013}
{Madiedo}, J.~M., {Trigo-Rodr{\'\i}guez}, J.~M., {Castro-Tirado}, A.~J.,
  {Ortiz}, J.~L., \& {Cabrera-Ca{\~n}o}, J. 2013, \mnras, 436, 2818

\bibitem[{{McMullan} {et~al.}(2023){McMullan}, {Vida}, {Devillepoix}, {Rowe},
  {Daly}, {King}, {Cup{\'a}k}, {Howie}, {Sansom}, {Shober}, {Towner},
  {Anderson}, {McFadden}, {Hor{\'a}k}, {Smedley}, {Joy}, {Shuttleworth},
  {Colas}, {Zanda}, {O'Brien}, {McMullan}, {Shaw}, {Suttle}, {Suttle}, {Young},
  {Campbell-Burns}, {Kacerek}, {Bassom}, {Bosley}, {Fleet}, {Jones},
  {McIntyre}, {James}, {Robson}, {Dickinson}, {Bland}, \&
  {Collins}}]{mcmullan2023}
{McMullan}, S., {Vida}, D., {Devillepoix}, H. A.~R., {et~al.} 2023, arXiv
  e-prints, arXiv:2303.12126

\bibitem[{{Nakano} \& {Hirabayashi}(2020)}]{nakano2020}
{Nakano}, R. \& {Hirabayashi}, M. 2020, \apjl, 892, L22

\bibitem[{{Neslu{\v{s}}an}(2015)}]{neslusan2015}
{Neslu{\v{s}}an}, L. 2015, Contributions of the Astronomical Observatory
  Skalnate Pleso, 45, 60

\bibitem[{{Ohtsuka} {et~al.}(2008){Ohtsuka}, {Arakida}, {Ito}, {Yoshikawa}, \&
  {Asher}}]{ohtsuka2008}
{Ohtsuka}, K., {Arakida}, H., {Ito}, T., {Yoshikawa}, M., \& {Asher}, D.~J.
  2008, Meteoritics and Planetary Science Supplement, 43, 5055

\bibitem[{{Ohtsuka} {et~al.}(2006){Ohtsuka}, {Sekiguchi}, {Kinoshita},
  {Watanabe}, {Ito}, {Arakida}, \& {Kasuga}}]{ohtsuka2006}
{Ohtsuka}, K., {Sekiguchi}, T., {Kinoshita}, D., {et~al.} 2006, \aap, 450, L25

\bibitem[{{Otto} {et~al.}(2023){Otto}, {Ho}, {Ulamec}, {Bibring}, {Biele},
  {Grott}, {Hamm}, {Hercik}, {Jaumann}, {Sato}, {Schr{\"o}der}, {Tanaka},
  {Auster}, {Kitazato}, {Knollenberg}, {Moussi}, {Nakamura}, {Okada},
  {Pilorget}, {Schmitz}, {Sugita}, {Wada}, \& {Yabuta}}]{otto2023}
{Otto}, K., {Ho}, T.-M., {Ulamec}, S., {et~al.} 2023, Earth, Planets and Space,
  75, 51

\bibitem[{{Picone} {et~al.}(2002){Picone}, {Hedin}, {Drob}, \&
  {Aikin}}]{picone2002}
{Picone}, J.~M., {Hedin}, A.~E., {Drob}, D.~P., \& {Aikin}, A.~C. 2002, Journal
  of Geophysical Research (Space Physics), 107, 1468

\bibitem[{{Plavec}(1950)}]{plavec1950}
{Plavec}, M. 1950, \nat, 165, 362

\bibitem[{{Prieto} {et~al.}(2013){Prieto}, {Madiedo}, {Trigo-Rodr{\'\i}guez},
  {Zamorano}, {Izquierdo}, {Oca{\~n}a}, {S{\'a}nchez de Miguel},
  {Castro-Tirado}, \& {Ortiz}}]{prieto2013}
{Prieto}, S., {Madiedo}, J.~M., {Trigo-Rodr{\'\i}guez}, J.~M., {et~al.} 2013,
  in 44th Annual Lunar and Planetary Science Conference, Lunar and Planetary
  Science Conference, 1108

\bibitem[{{Richardson} {et~al.}(1998){Richardson}, {Bottke}, \&
  {Love}}]{richardson1998}
{Richardson}, D.~C., {Bottke}, W.~F., \& {Love}, S.~G. 1998, \icarus, 134, 47

\bibitem[{{Robertson} \& {Mathias}(2017)}]{robertson2017}
{Robertson}, D.~K. \& {Mathias}, D.~L. 2017, Journal of Geophysical Research
  (Planets), 122, 599

\bibitem[{{Ryabova}(2007)}]{ryabova2007}
{Ryabova}, G.~O. 2007, \mnras, 375, 1371

\bibitem[{{Ryabova}(2016)}]{ryabova2016}
{Ryabova}, G.~O. 2016, \mnras, 456, 78

\bibitem[{{Ryabova} {et~al.}(2019){Ryabova}, {Avdyushev}, \&
  {Williams}}]{ryabova2019}
{Ryabova}, G.~O., {Avdyushev}, V.~A., \& {Williams}, I.~P. 2019, \mnras, 485,
  3378

\bibitem[{{Shuvalov} \& {Artemieva}(2002)}]{shuvalov2002}
{Shuvalov}, V.~V. \& {Artemieva}, N.~A. 2002, \planss, 50, 181

\bibitem[{{Slyuta}(2017)}]{slyuta2017}
{Slyuta}, E.~N. 2017, Solar System Research, 51, 64

\bibitem[{{Spurn{\'y}}(1993)}]{spurny1993}
{Spurn{\'y}}, P. 1993, in Meteoroids and their Parent Bodies, ed. J.~{Stohl} \&
  I.~P. {Williams}, 193

\bibitem[{{Spurn{\'y}} {et~al.}(2012){Spurn{\'y}}, {Bland}, {Shrben{\'y}},
  {Borovi{\v{c}}ka}, {Ceplecha}, {Singelton}, {Bevan}, {Vaughan}, {Towner},
  {McClafferty}, {Toumi}, \& {Deacon}}]{spurny2012}
{Spurn{\'y}}, P., {Bland}, P.~A., {Shrben{\'y}}, L., {et~al.} 2012, \maps, 47,
  163

\bibitem[{{Spurn{\'y}} {et~al.}(2017){Spurn{\'y}}, {Borovi{\v{c}}ka},
  {Baumgarten}, {Haack}, {Heinlein}, \& {S{\o}rensen}}]{spurny2017}
{Spurn{\'y}}, P., {Borovi{\v{c}}ka}, J., {Baumgarten}, G., {et~al.} 2017,
  \planss, 143, 192

\bibitem[{{Spurn{\'y}} {et~al.}(2007){Spurn{\'y}}, {Borovi{\v{c}}ka}, \&
  {Shrben{\'y}}}]{spurny2007}
{Spurn{\'y}}, P., {Borovi{\v{c}}ka}, J., \& {Shrben{\'y}}, L. 2007, in Near
  Earth Objects, our Celestial Neighbors: Opportunity and Risk, ed. G.~B.
  {Valsecchi}, D.~{Vokrouhlick{\'y}}, \& A.~{Milani}, Vol. 236, 121--130

\bibitem[{{Svetsov} {et~al.}(1995){Svetsov}, {Nemtchinov}, \&
  {Teterev}}]{svetsov1995}
{Svetsov}, V.~V., {Nemtchinov}, I.~V., \& {Teterev}, A.~V. 1995, \icarus, 116,
  131

\bibitem[{{Todorovi{\'c}}(2018)}]{todorovic2018}
{Todorovi{\'c}}, N. 2018, \mnras, 475, 601

\bibitem[{{Trigo-Rodr{\'\i}guez} \& {Llorca}(2006)}]{trigo2006}
{Trigo-Rodr{\'\i}guez}, J.~M. \& {Llorca}, J. 2006, \mnras, 372, 655

\bibitem[{{Tsuchiyama} {et~al.}(2008){Tsuchiyama}, {Mashio}, {Imai}, {Noguchi},
  {Miura}, \& {Yano}}]{tsuchiyama2008}
{Tsuchiyama}, A., {Mashio}, E., {Imai}, Y., {et~al.} 2008, in Procedings of
  Japan Geoscience Union Meeting, Hakodate, Japan

\bibitem[{{{\v{C}}apek} \& {Borovi{\v{c}}ka}(2009)}]{capek2009}
{{\v{C}}apek}, D. \& {Borovi{\v{c}}ka}, J. 2009, \icarus, 202, 361

\bibitem[{{{\v{C}}apek} \& {Vokrouhlick{\'y}}(2010)}]{capek2010}
{{\v{C}}apek}, D. \& {Vokrouhlick{\'y}}, D. 2010, \aap, 519, A75

\bibitem[{{{\v{C}}apek} \& {Vokrouhlick{\'y}}(2011)}]{capek2011}
{{\v{C}}apek}, D. \& {Vokrouhlick{\'y}}, D. 2011, in EPSC-DPS Joint Meeting
  2011, Vol. 2011, 867

\bibitem[{{{\v{C}}apek} \&
  {Vokrouhlick{\'y}}(2012{\natexlab{a}})}]{capek2012lpi}
{{\v{C}}apek}, D. \& {Vokrouhlick{\'y}}, D. 2012{\natexlab{a}}, in LPI
  Contributions, Vol. 1667, Asteroids, Comets, Meteors 2012, ed. {LPI Editorial
  Board}, 6051

\bibitem[{{{\v{C}}apek} \& {Vokrouhlick{\'y}}(2012{\natexlab{b}})}]{capek2012}
{{\v{C}}apek}, D. \& {Vokrouhlick{\'y}}, D. 2012{\natexlab{b}}, \aap, 539, A25

\bibitem[{{Voj{\'a}{\v{c}}ek} {et~al.}(2015){Voj{\'a}{\v{c}}ek},
  {Borovi{\v{c}}ka}, {Koten}, {Spurn{\'y}}, \& {{\v{S}}tork}}]{vojacek2015}
{Voj{\'a}{\v{c}}ek}, V., {Borovi{\v{c}}ka}, J., {Koten}, P., {Spurn{\'y}}, P.,
  \& {{\v{S}}tork}, R. 2015, \aap, 580, A67

\bibitem[{{Weibull}(1939)}]{weibull1939}
{Weibull}, W. 1939, {A Statistical Theory of the Strength of Materials}
  (Stockholm, Sweden: Generalstabens Litografiska Anstalts F{\"o}rlag)

\bibitem[{{Weibull}(1951)}]{weibull1951}
{Weibull}, W. 1951, Journal of Applied Mechanics, 18, 293

\bibitem[{{Whipple}(1983)}]{whipple1983}
{Whipple}, F.~L. 1983, \iaucirc, 3881, 1

\bibitem[{{Williams} \& {Wu}(1993)}]{williams1993}
{Williams}, I.~P. \& {Wu}, Z. 1993, \mnras, 262, 231

\bibitem[{{Yada} {et~al.}(2021){Yada}, {Abe}, {Okada}, {Nakato}, {Yogata},
  {Miyazaki}, {Hatakeda}, {Kumagai}, {Nishimura}, {Hitomi}, {Soejima},
  {Yoshitake}, {Iwamae}, {Furuya}, {Uesugi}, {Karouji}, {Usui}, {Hayashi},
  {Yamamoto}, {Fukai}, {Sugita}, {Cho}, {Yumoto}, {Yabe}, {Bibring},
  {Pilorget}, {Hamm}, {Brunetto}, {Riu}, {Lourit}, {Loizeau}, {Lequertier},
  {Moussi-Soffys}, {Tachibana}, {Sawada}, {Okazaki}, {Takano}, {Sakamoto},
  {Miura}, {Yano}, {Ireland}, {Yamada}, {Fujimoto}, {Kitazato}, {Namiki},
  {Arakawa}, {Hirata}, {Yurimoto}, {Nakamura}, {Noguchi}, {Yabuta}, {Naraoka},
  {Ito}, {Nakamura}, {Uesugi}, {Kobayashi}, {Michikami}, {Kikuchi}, {Hirata},
  {Ishihara}, {Matsumoto}, {Noda}, {Noguchi}, {Shimaki}, {Shirai}, {Ogawa},
  {Wada}, {Senshu}, {Yamamoto}, {Morota}, {Honda}, {Honda}, {Yokota},
  {Matsuoka}, {Sakatani}, {Tatsumi}, {Miura}, {Yamada}, {Fujii}, {Hirose},
  {Hosoda}, {Ikeda}, {Iwata}, {Kikuchi}, {Mimasu}, {Mori}, {Ogawa}, {Ono},
  {Shimada}, {Soldini}, {Takahashi}, {Takei}, {Takeuchi}, {Tsukizaki},
  {Yoshikawa}, {Terui}, {Nakazawa}, {Tanaka}, {Saiki}, {Yoshikawa}, {Watanabe},
  \& {Tsuda}}]{yada2022}
{Yada}, T., {Abe}, M., {Okada}, T., {et~al.} 2021, Nature Astronomy, 6, 214

\bibitem[{{Yu} {et~al.}(2019){Yu}, {Ip}, \& {Spohn}}]{yu2019}
{Yu}, L.~L., {Ip}, W.~H., \& {Spohn}, T. 2019, \mnras, 482, 4243

\bibitem[{{Zhang} {et~al.}(2023){Zhang}, {Battams}, {Ye}, {Knight}, \&
  {Schmidt}}]{zhang2023}
{Zhang}, Q., {Battams}, K., {Ye}, Q., {Knight}, M.~M., \& {Schmidt}, C.~A.
  2023, \psj, 4, 70

\bibitem[{{Zhdan} {et~al.}(2007){Zhdan}, {Stulov}, {Stulov}, \&
  {Turchak}}]{zhdan2007}
{Zhdan}, I.~A., {Stulov}, V.~P., {Stulov}, P.~V., \& {Turchak}, L.~I. 2007,
  Solar System Research, 41, 505

\end{thebibliography}
\end{document}